\documentclass[acmtog,authorversion]{acmart}

\AtBeginDocument{%
  }

\usepackage{lscape}
\usepackage{listings}

\lstdefinestyle{promptstyle}{
    basicstyle=\ttfamily\footnotesize,
    breaklines=true,
    breakindent=0pt,
    frame=single,
    backgroundcolor=\color{gray!10},
    showstringspaces=false,
    columns=fullflexible,
    keepspaces=true,
}

\copyrightyear{2025} 
\acmYear{2025} 
\setcopyright{cc}
\setcctype{by}
\acmConference[SIGGRAPH Conference Papers '25]{Special Interest Group on Computer Graphics and Interactive Techniques Conference Conference Papers}{August 10--14, 2025}{Vancouver, BC, Canada}
\acmBooktitle{Special Interest Group on Computer Graphics and Interactive Techniques Conference Conference Papers (SIGGRAPH Conference Papers '25), August 10--14, 2025, Vancouver, BC, Canada}
\acmDOI{10.1145/3721238.3730762}
\acmISBN{979-8-4007-1540-2/2025/08}

\begin{document}

\title{A Platform for Interactive AI Character Experiences}

\author{Rafael Wampfler}
\affiliation{%
\institution{ETH Zurich}
\city{Zurich}
\country{Switzerland}
}
\email{rafael.wampfler@inf.ethz.ch}

\author{Chen Yang}
\affiliation{%
\institution{ETH Zurich}
\city{Zurich}
\country{Switzerland}
}
\email{chen.yang@inf.ethz.ch}

\author{Dillon Elste}
\affiliation{%
\institution{ETH Zurich}
\city{Zurich}
\country{Switzerland}
}
\email{delste@inf.ethz.ch}

\author{Nikola Kova\v{c}evi\'{c}}
\affiliation{%
\institution{ETH Zurich}
\city{Zurich}
\country{Switzerland}
}
\email{nikola.kovacevic@inf.ethz.ch}

\author{Philine Witzig}
\affiliation{%
\institution{ETH Zurich}
\city{Zurich}
\country{Switzerland}
}
\email{philine.witzig@inf.ethz.ch}

\author{Markus Gross}
\affiliation{%
\institution{ETH Zurich}
\city{Zurich}
\country{Switzerland}
}
\email{grossm@inf.ethz.ch}

\renewcommand{\shortauthors}{Wampfler, et al.}

\begin{abstract}

From movie characters to modern science fiction --- bringing characters into interactive, story-driven conversations has captured imaginations across generations. Achieving this vision is highly challenging and requires much more than just language modeling. It involves numerous complex AI challenges, such as conversational AI, maintaining character integrity, managing personality and emotions, handling knowledge and memory, synthesizing voice, generating animations, enabling real-world interactions, and integration with physical environments.  Recent advancements in the development of foundation models, prompt engineering, and fine-tuning for downstream tasks have enabled researchers to address these individual challenges. However, combining these technologies for interactive characters remains an open problem.
We present a system and platform for conveniently designing believable digital characters, enabling a conversational and story-driven experience while providing solutions to all of the technical challenges. As a proof-of-concept, we introduce \emph{Digital Einstein}, which allows users to engage in conversations with a digital representation of Albert Einstein about his life, research, and persona. While \emph{Digital Einstein} exemplifies our methods for a specific character, our system is flexible and generalizes to any story-driven or conversational character. By unifying these diverse AI components into a single, easy-to-adapt platform, our work paves the way for immersive character experiences, turning the dream of lifelike, story-based interactions into a reality.

\end{abstract}



\begin{teaserfigure}
    \centering
    \includegraphics[width=\textwidth]{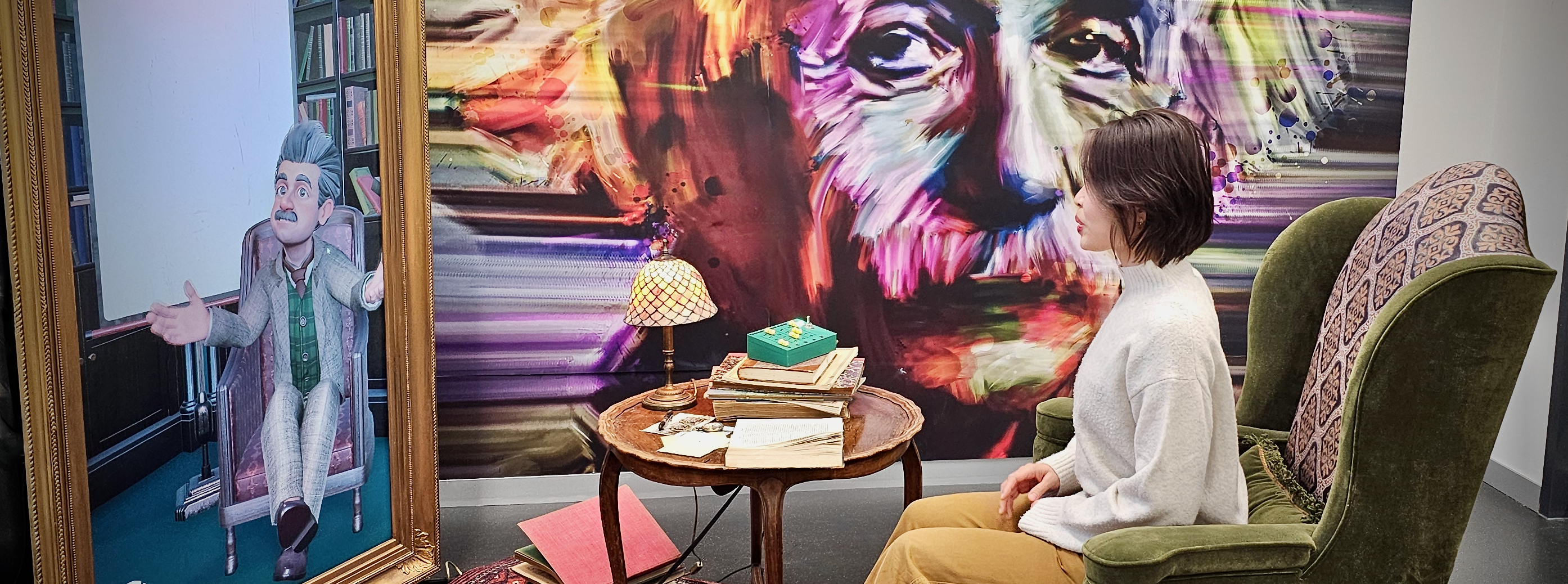}
    \caption{Our system enables conversational and story-driven experiences with a digital character. Users can engage with the character on any topic through speech while the character mimics human behavior naturally.}
    \label{fig:teaser}
    \Description[Digital Einstein character in interactive setup]{A digital character of Einstein is shown in a lifelike setup, interacting with a user. The character mimics human behavior, responding to speech input with expressive animations. The environment is furnished to match an early 20th-century setup, enhancing immersion.}
\end{teaserfigure}

\maketitle

\section{Introduction}

The vision of creating interactive, lifelike digital characters capable of engaging in meaningful, story-driven conversations has fascinated generations~\cite{cavazza2002interacting, bates1994role, lugrin2022handbook}. 
From movie characters to digital representations of historical figures --- such characters redefine how we experience storytelling and establish emotional connections to digital entities~\cite{torre2019effect, xu2024filmagent, gong2023interactive}. 

However, realizing this vision is a challenging task. It requires a seamless combination of conversational intelligence~\cite{ramesh2017survey}, character integrity~\cite{schlenker2008integrity}, personality and emotion~\cite{bates1994role}, knowledge and memory~\cite{kope2013modeling}, voice synthesis~\cite{torre2019effect}, realistic animations~\cite{wu2024probtalk3d}, and integration into the physical environment~\cite{li2022interactive}. Due to this complexity, actors and actresses, as in Disney's "Turtle Talk with Crush", puppeteered characters in real-time~\cite{casas2019intermediated}. Even with major advancements in AI technology, conversational systems still struggle to provide interactive and story-driven experiences~\cite{riedl2003managing,green2014interactive}. Strategies often focus on isolated components, such as animation synthesis~\cite{ng2024audio, liu2022beat, kim2024body} or language modeling~\cite{schmitt2021characterchat}, rather than ensuring overall coherence across all components. Furthermore, requirements such as character consistency, customization, and real-time synchronization of the components often fall short~\cite{lugrin2022handbook}. 

In this work, we propose a modular system for creating believable conversational digital characters that support narrative experiences. 
By combining the power of large language models (LLMs) with multimodal sensing, expressive synthesis, and adaptive personality modeling, this system addresses the pertaining challenges, allowing for interactive, story-driven, and believable interactions.
As a proof-of-concept, we present \emph{Digital Einstein}, a digital representation of Albert Einstein, enabling users to have discussions on his scientific research, anecdotes from his life, and historical background. 
The system creates an immersive experience by integrating a story-driven AI character into a physical environment (see Figure~\ref{fig:teaser}). 
While \emph{Digital Einstein} only serves as an example application, the system architecture is highly modular.
Individual components can be easily exchanged depending on the character and the specific target application.
Thus, our work opens new possibilities for bringing interactive and believable digital characters to life.

Our system contributes several technical innovations that enable believable digital characters. It maintains character integrity using GPT-4o and a fine-tuned Llama 3 model, enhanced by synthetic conversation generation, embedding-based prompt steering, and a memory system for story consistency. Personality is dynamically adjustable, with emotional tone expressed through both speech and animation. Further, conversations are visually enriched with images generated by Midjourney. In addition, the character interprets its physical environment through a camera, enabling situational awareness. These components are integrated into a modular and extensible platform suited for diverse conversational and story-driven applications, anchored in a themed physical setup for immersive and emotionally engaging interactions.
\section{Related Work}

\subsection{Conversational Digital Characters}

\begin{figure*}[tb]
    \centering
    \includegraphics[width=0.98\linewidth,trim={0cm 10.2cm 7.6cm 0cm},clip]{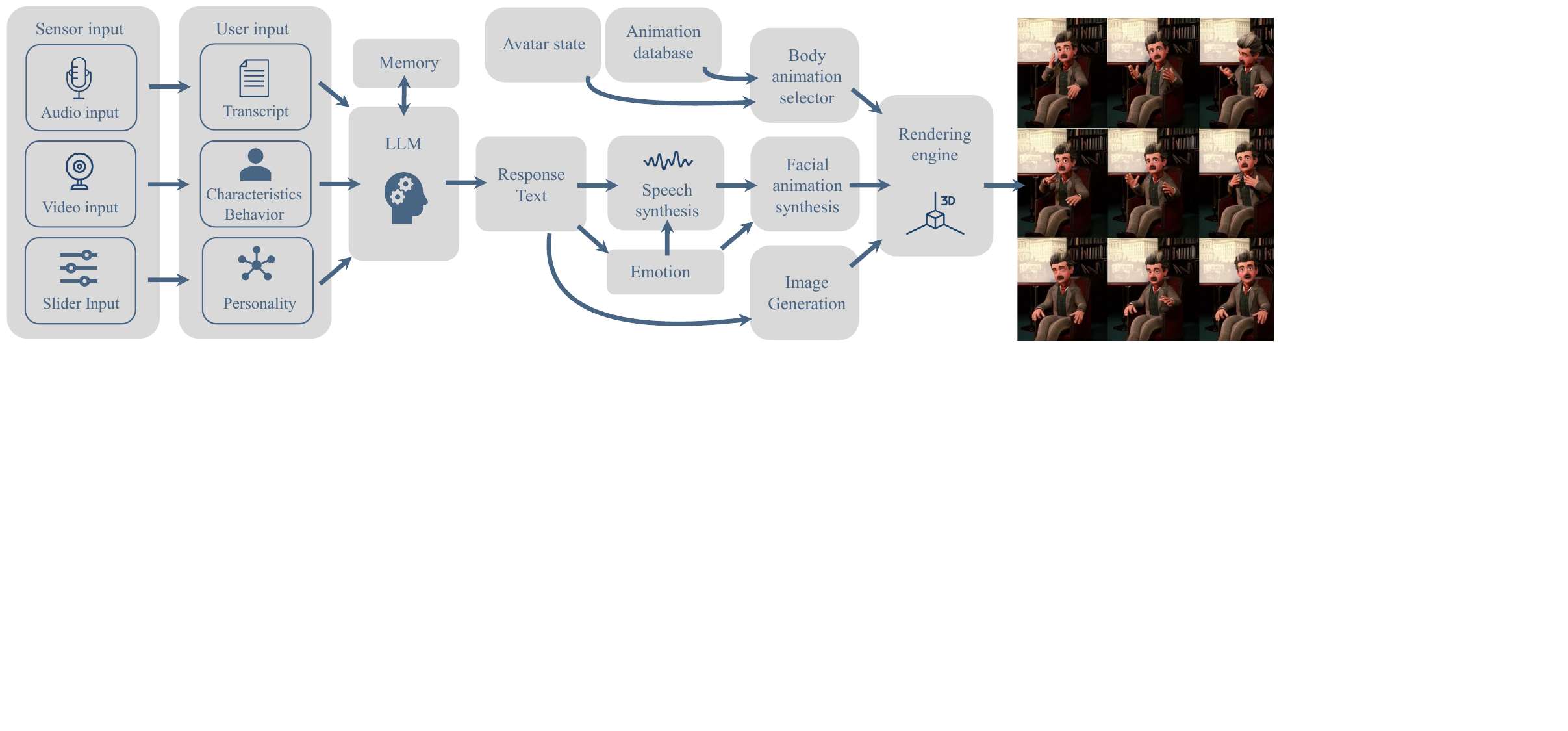}
    \caption{System Overview: The pipeline processes sensor inputs, including transcribed speech and video-based user characteristics and behavior analysis, through an LLM-based chatbot supported by memory and an adjustable personality of the digital character. The chatbot's responses guide speech and facial animation synthesis based on emotions detected in the response, motion-captured body animation selected based on the avatar state, and image generation.}
    \label{fig:pipeline}
    \Description[System architecture overview of AI character pipeline]{A modular system pipeline diagram showing the flow from sensor input (audio, video, personality sliders) to response generation via LLMs, emotion recognition, voice synthesis, animation selection, and rendering. Modules are separated, emphasizing the integration of multimodal inputs to drive digital character behavior.}
\end{figure*}

Conversational digital characters have progressed from rule-based systems~\cite{mateas2003integrating} to LLM-powered models~\cite{qi2021conversational}, enabling dynamic, context-rich conversations. Personality modeling advances include dynamic personality infusion, where chatbot responses reflect predefined traits~\cite{kovacevic2024personality,kovacevic2024chatbots}. Emotion-aware systems improve user engagement through speech emotion recognition~\cite{hu2022acoustically} and text-based emotion detection~\cite{kusal2024understanding}. Further, modular architectures promote conversation consistency and scalability~\cite{nguyen2022exploring}.

Conversational AI finds applications in education, healthcare, and storytelling. Narrative agents foster engagement through authored dialogue~\cite{spierling2005beyond}, while LLM-based storytelling supports coherent narratives~\cite{li2024words}. These systems simplify complex tasks and enhance accessibility~\cite{qi2021conversational}. However, sustaining meaningful interaction over multiple exchanges remains a core challenge. In particular, addressing memory limitations is key to maintaining multiturn coherence~\cite{castillo2024beyond,johri2025evaluation}. To overcome these memory limitations, RAG combines retrieval and generation to help chatbots maintain long-term context~\cite{gao2023retrieval}. Dual-memory systems balance short- and long-term data for personalization~\cite{zhang2024advancing} and flow, while selective memory improves user experience and retrieval efficiency~\cite{sumida2024should}.

\subsection{AI-Driven Narratives with Ethical Considerations}

Early approaches, such as a semi-automatic artistic pipeline for recreating Einstein, demonstrated how small-scale productions could achieve realism with limited resources~\cite{helzle2018digital}. 
Building on this, AI-driven conversational agents, as seen in the "Living Memories" concept, brought figures like Leonardo da Vinci to life~\cite{pataranutaporn2023living}. 
Larger-scale efforts, such as developing corpora for role-playing Chinese historical figures, highlighted the importance of contextual authenticity and low-resource data integration for nuanced depictions~\cite{bai2024baijia}. 
Meanwhile, ethical considerations have gained prominence. 
Research in "digital necromancy" examined the balance between preserving cultural heritage and addressing issues of authenticity and consent~\cite{hutson2023life}. Recent work shows how LLMs improve accessibility in digital humanities by generating concise portrayals of historical figures~\cite{hasnain2024potential}, while ethical frameworks guide the reconstruction of narratives in education~\cite{hutson2024digital}.

\subsection{Interactive Systems}

Interactive systems have advanced conversational characters, enabling lifelike and engaging interactions. Recent frameworks align body movements with co-speech gestures, producing emotionally rich and context-aware responses~\cite{kim2024body}. Modular architectures further support such interactions by decoupling dialog management from embodiment, enabling customization and robust nonverbal communication~\cite{santos2023cube}. Other end-to-end pipelines enhance virtual agents with real-time audio-video synchronization and anthropomorphic features~\cite{rupprecht2024digital}.

Immersive augmented and mixed reality systems also benefit from these advances. Systems combining speech recognition with real-time facial animation enhance character interactions in augmented reality~\cite{casas2023intermediated}. MoodFlow~\cite{casas2024moodflow} extends this by using a prompt-embedded state machine to guide emotionally intelligent avatars in mixed reality. Platforms integrating vision and language models enable context-aware, real-time interactions~\cite{maniatis2023voxreality}, while hybrid systems support seamless user experiences through unobtrusive, spatially immersive interfaces~\cite{encarnacao2000walk}.

Unlike previous work, we address the AI challenges of story-driven and interactive conversations with believable characters through a unified, customizable framework.
\section{System Design and Requirements}

\subsection{System Requirements}

The development of our system was guided by several goals (i.e., believability, flexibility, realism) and constraints (i.e., low latency, resilience, technical complexity) that define our system requirements:

\begin{enumerate}
    \item \textit{Modular and Scalable Design:} Supports easy upgrades and adaptations for diverse contexts.

    \item \textit{User-Centric Approach:} Ensures intuitive and customizable interactions, allowing users to adjust the character's personality and tailor conversations to their preferences.

    \item \textit{Real-Time Responsiveness:} Maintains low latency across all components to ensure seamless dialogue.

    \item \textit{Robustness and Reliability:} Guarantees smooth operation in different settings, even under varying conditions.

    \item \textit{Immersive Experience:} Combines a themed physical setup with spatial audio, realistic animations, and lifelike body movements synchronized with AI-generated responses to ensure natural and engaging interactions.
\end{enumerate}

\subsection{System Overview}

Our system comprises several interconnected AI modules to address the complex AI challenges for interactive, story-driven characters. Figure~\ref{fig:pipeline} provides a high-level overview of the connection between our modules. A detailed interaction flow is shown in Figure~\ref{fig:uml}.

The system's core component, implemented in Unity, features a digital character within a themed scene. 
The character transitions between four distinct states: \emph{idle} (no interaction), \emph{listening} (awaiting user input), \emph{thinking} (processing input), and \emph{speaking} (delivering responses). 
Inviting animations are played when a user is approaching, which is detected by the camera (our system queries the camera every $500$ms). 
When the user sits down, the character transitions from \emph{idle} to \emph{speaking}, starting the conversation with a randomly selected welcome message, followed by awaiting user input. 
While \emph{thinking}, the transcribed user input is processed by the cognitive module supported by an LLM-based chatbot. 
The system ensures character integrity by maintaining consistency in behavior and dialogue. 
Additionally, the character draws on its knowledge base and memory to provide contextually relevant responses. 
Users can further personalize their experience by adjusting the character's personality traits using physical sliders that dynamically affect response patterns. 
From the chatbot response, emotions are extracted to adjust speech and animations. 
Speech is synthesized using a fine-tuned Microsoft Azure neural voice model. 
The character's animations blend facial expressions, dynamically generated and synchronized from speech using Audio2Face~\cite{karras2017audio}, with motion-captured body movements. 
While the character is \emph{speaking}, images are automatically generated using Midjourney based on the conversation context.
Throughout the interaction, the character alternates between \emph{listening}, \emph{thinking}, and \emph{speaking}. 
A fluid dialogue flow is maintained through coordinated state transitions based on user behavior and system responses. 
If the user remains silent for more than seven seconds, a signal is sent to the chatbot, prompting it to respond appropriately.
\section{Design Challenges and Solutions} \label{chapter:challenge}

In this section, we discuss the key challenges encountered in developing a system for believable, conversational, and story-driven characters, along with our solutions. 

\subsection{Conversational Intelligent Chatbots}
Creating believable digital characters requires sophisticated conversational abilities. 
The core challenge is to develop chatbots that can process natural language, generate contextually appropriate responses, and maintain coherent, story-driven interactions that support user immersion.
To address this challenge, we leverage state-of-the-art LLMs, specifically GPT-4o~\cite{achiam2023gpt} and Llama 3 8B~\cite{touvron2023llama}. 
GPT-4o is used for real-time thematic consistency and high-quality engagement, while Llama 3 supports local deployment scenarios requiring privacy and cost efficiency, making it a valuable offline alternative. This dual-model setup enhances robustness, real-time responsiveness, and ensures continuity during cloud service outages, addressing requirements (3) and (4).

\begin{figure}[tb]
    \centering
    \includegraphics[width=\columnwidth,trim={0.7cm 9.7cm 19.7cm 0.5cm},clip]{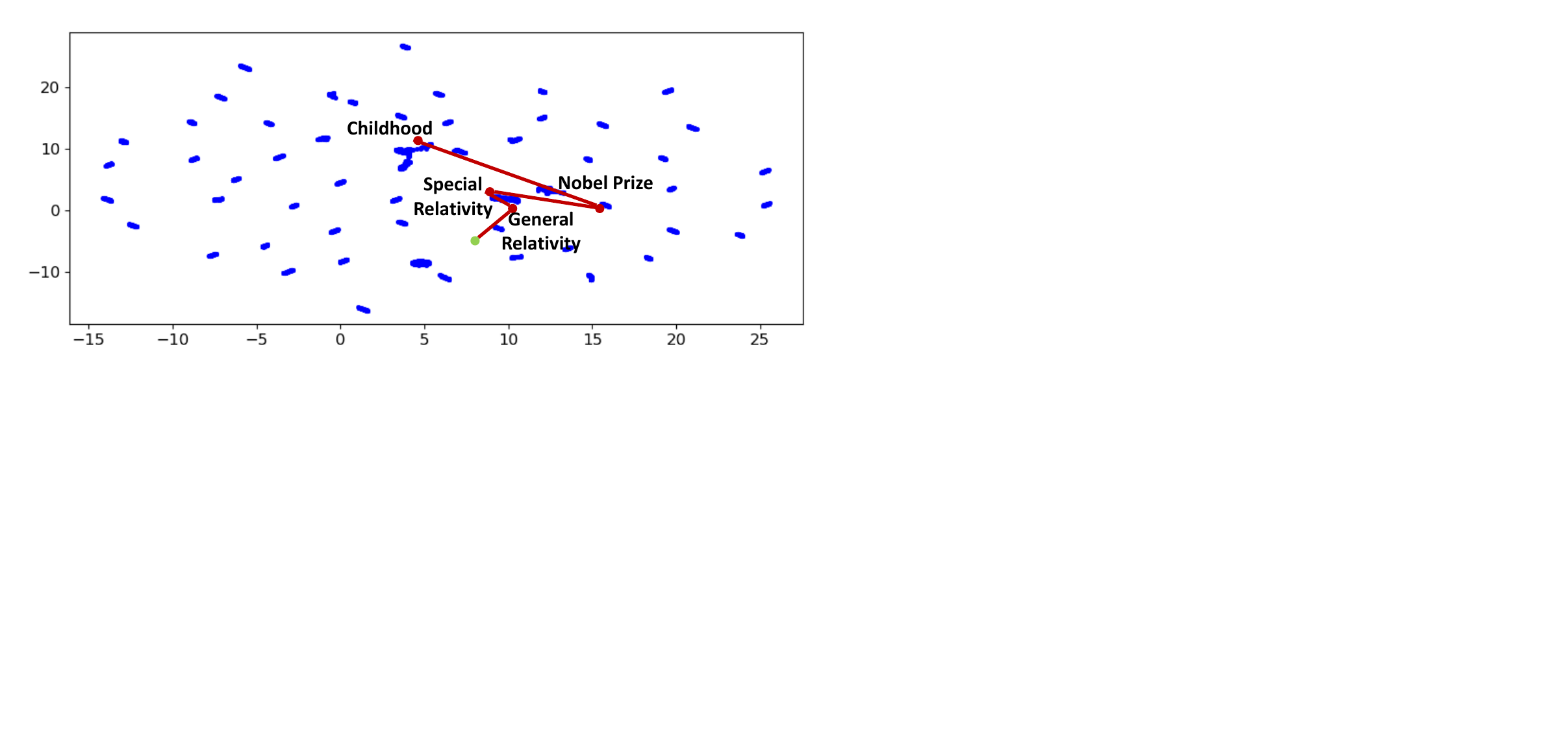}
    \caption{Visualization of the embedding space from synthetic Einstein conversations. The clusters of blue points represent different topics. A user interaction trace is highlighted in red, with the starting point marked in green. The transition from the topic "Nobel Prize" to "childhood" is initiated by the user.}
    \label{fig:embeddings}
    \Description[Embedding space visualization of Einstein conversations]{A 2D UMAP projection of high-dimensional embeddings representing synthetic Einstein conversations across various topics. Distinct clusters indicate topics like 'Nobel Prize' and 'Childhood'. A red trajectory line with a green starting point visualizes a user's conversational path moving between these topics.}
\end{figure}

\paragraph{Character Stories and Topics}
We compiled $M$ topics related to the character's expertise and personal interests, such as hobbies and anecdotes. 
For the Einstein character, this set includes $M=62$ topics in total, spanning his personal life, scientific theories, and musical interests.
We used GPT-4o to generate $N$ synthetic human-character conversations for each topic (see supplemental material). 
Each conversation turn $t$ was embedded using Microsoft Azure's \textit{text-embedding-3-large} model, producing a 3072-dimensional vector $\mathbf{e}_t$.
To compute a representative vector for each conversation, we averaged its turn-level embeddings: $\mathbf{e}_{\text{conv}} = \frac{1}{T} \sum_{t=1}^{T} \mathbf{e}_t$, where $T$ is the number of turns in the conversation. To obtain a topic-level representation, we further averaged all conversation embeddings within that topic: $\mathbf{e}_{\text{topic}}^{(j)} = \frac{1}{N} \sum_{i=1}^{N} \mathbf{e}_{\text{conv}}^{(i,j)}, j = 1, \dots, M$.
Figure~\ref{fig:embeddings} shows a 2D UMAP projection where each point corresponds to $\mathbf{e}_{\text{conv}}$. 
Distinct clusters can be found per topic, and thematically related topics are positioned close to each other. 
The red trace illustrates a user's multi-topic conversation path through the embedding space.

\paragraph{Large Language Models}
We fine-tuned Llama 3 8B to enhance its topic consistency and knowledge of character-specific topics on the raw text of the synthetic conversations. 
We used Axolotl~\cite{axolotl2025} and fine-tuned for three epochs with the Adam optimizer (learning rate of $2e^{-5}$, gradient accumulation of 8, cosine scheduling with 100 warmup steps) using a batch size of 1 on a cloud-based NVIDIA RTX 6000 GPU. 
For efficient operation, addressing requirement (3), Llama 3 8B runs locally with DeepSpeed~\cite{rasley2020deepspeed}.

For GPT-4o (Microsoft Azure, version 2024-08-06), we use prompting to induce topic consistency. 
The system tracks the conversation flow using the embedding space. Each new turn is embedded and combined with past turns of the same session using an exponentially weighted average with a three-turn half-life ($h=3$), giving higher weight to recent turns: $\mathbf{e}_{\text{agg}} = \sum_{k=0}^{\infty} \alpha_k \, \mathbf{e}_{t-k}$, where $\alpha_k = \frac{1}{Z} \cdot 2^{-k/h}$ and $Z = \frac{1}{1 - 2^{-1/h}}$. Here, $\mathbf{e}_{t-k}$ is the embedding of the turn $k$ steps ago. 
The system then computes the cosine similarity between this aggregated embedding and all topic embeddings $\mathbf{e}_{\text{topic}}^{(j)}, j = 1, \dots, M$. If the highest similarity exceeds a defined threshold, the corresponding topic $j$ is selected as the \textit{current topic}. To facilitate smooth topic transitions, the system performs a neighbor search between $\mathbf{e}_{\text{agg}}$ and all topic embeddings $\mathbf{e}_{\text{topic}}^{(j)}, j = 1, \dots, M$. One of the three nearest neighbors is randomly selected as the \textit{next topic}.
The prompt is then modified to steer GPT-4o toward a smooth transition to the \textit{next topic}.

For both models, the prompt first defines the role of the digital character and then six distinct contexts, followed by instructions and the conversation history (see Table~\ref{tab:chatbot_prompt} and supplemental material).

\renewcommand{\arraystretch}{1.6}
\begin{table}[tb]
\centering
\caption{Element descriptions and example responses for chatbot prompts.}
\label{tab:chatbot_prompt}
\Description[Chatbot prompt elements and sample responses]{A table listing key prompt components used to steer LLM-based chatbot behavior. Includes fields like Location, Scene, User Description, Date, Image Description, Memory, and Instructions, along with example responses to guide character context.}
\resizebox{\columnwidth}{!}{%
\begin{tabular}{l|p{9cm}}
    Prompt Element & Description \& Example Response \\ \hline \hline
    Location & Current location \newline 
    \textit{Vancouver's vibrant energy at SIGGRAPH is a perfect reflection of the curious minds gathered here to shape the future of innovation.}
    \\ \hline
    Scene &  Descriptions of physical setup and Unity scene \newline
    \textit{This cozy setting feels perfect for deep, thoughtful conversations, as if we're sharing ideas over a cup of tea in a timeless library.}
    \\ \hline
    User Description & Age, gender, number of people around, user appearance, user attention \newline
    \textit{You look ready for adventure in your blue shorts and white shirt, perfect for a curious mind like yours!}
     \\ \hline
     Image Description & Metadata of displayed image \newline
     \textit{The swirling star remnants and slowing clocks perfectly capture how a black hole warps light and time, truly a cosmic wonder!}
     \\ \hline
     Date & Day and time \newline
     \textit{Ah, nearly midnight, a wonderfully quiet time when the mind can wander freely and explore its most curious thoughts!}
     \\ \hline
     Memory & 5 Turns from past conversations
     \\ \hline
     Instructions &  Persona, tone, response guidelines, behavior, handling specific situations (e.g., silence), topic transitions.
     \\ \hline
    Additional Data & Current conversation history \\ \hline
\end{tabular}
}
\end{table}

\paragraph{Knowledge and Memory} 
Our system implements a vector-based knowledge store to maintain conversation history.
For each new user input, the system generates an embedding and performs a similarity search against stored conversations. 
The five most relevant conversation snippets are incorporated into the prompt's memory context. 
By focusing on topics and information most relevant to the current conversation context, this approach fulfills requirement (3).

\subsection{Personality and Emotion}
\label{sec:personality_emotion}
Beyond basic conversational abilities, a compelling digital character must master multiple dimensions of human-like interaction. 
Key among these is maintaining consistent personality traits while adapting to different conversational contexts, and demonstrating emotional intelligence through appropriate responses.. 
To meet requirement (5), we integrated emotional intelligence into our system.

An LLM can dynamically adapt the emotional tone of the responses based on the user's input~\cite{chang2024behavioral,jin2024implicit}. 
Thus, after response generation, we use GPT-4o-mini to determine one of 7 emotions matching the emotions supported by Audio2Face (i.e., amazement, anger, disgust, fear, joy, sadness, neutral). 
Furthermore, an intensity level ranging from 0.01 to 2.0 is regressed. 
These parameters are passed to Microsoft Azure Speech Synthesis, which adjusts its speaking style using the following mapping: amazement $\rightarrow$ excited, anger $\rightarrow$ angry, disgust $\rightarrow$ disgruntled, fear $\rightarrow$ fearful, joy $\rightarrow$ cheerful, sadness $\rightarrow$ sad, neutral $\rightarrow$ default style.

\begin{figure}[tb]
    \centering
    \includegraphics[width=\columnwidth,trim={14cm 19cm 0cm 29cm},clip]{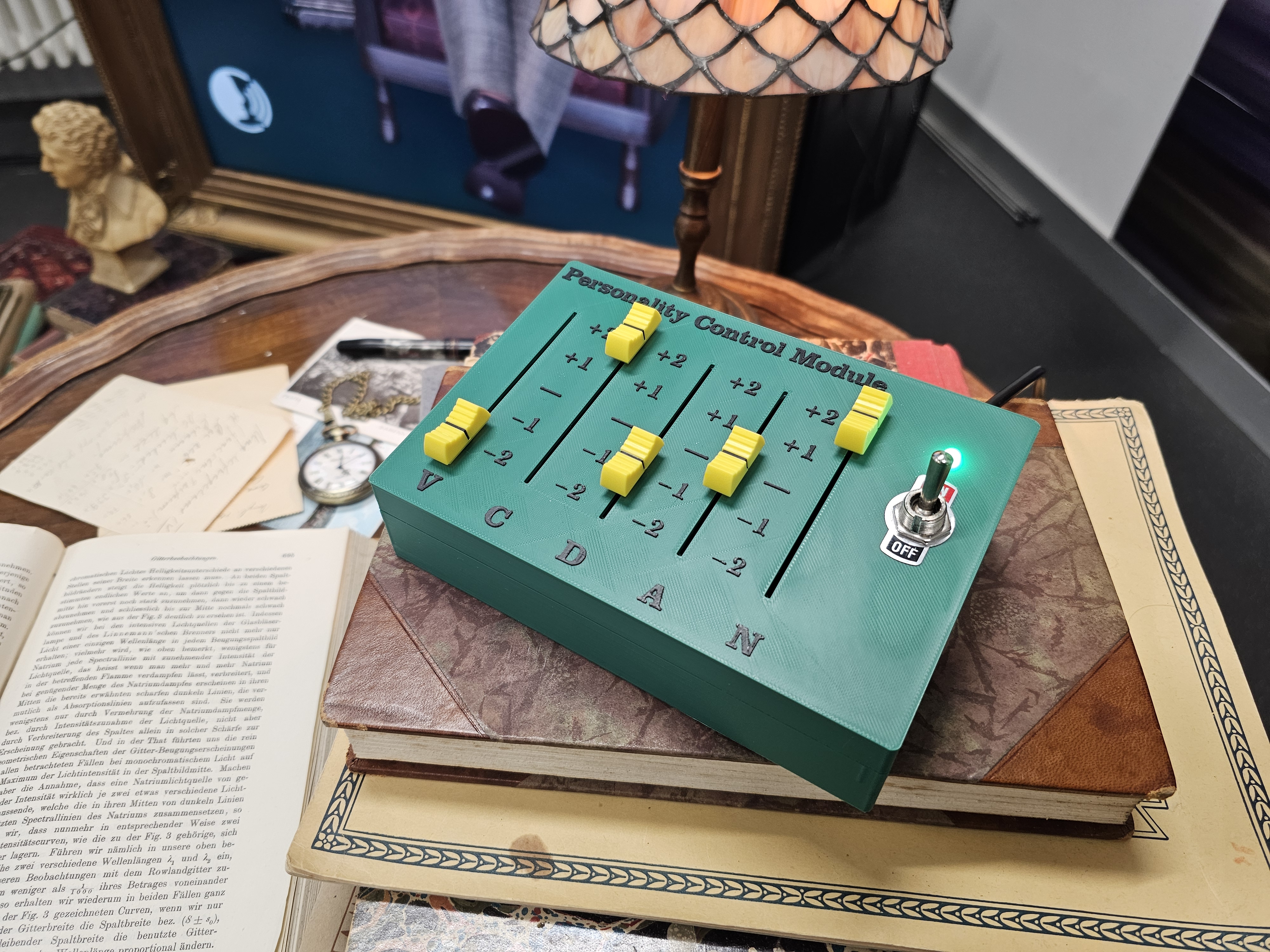}
    \caption{Sliders for real-time personality adjustment across five traits: \textbf{V}ibrancy, \textbf{C}onscientiousness, \textbf{D}ecency, \textbf{A}rtificiality, and \textbf{N}euroticism.}
    \label{fig:sliders}
    \Description[Physical sliders for personality control]{Five labeled physical sliders allow users to adjust character personality traits: Vibrancy, Conscientiousness, Decency, Artificiality, and Neuroticism. Each slider is mapped to a corresponding trait value that dynamically alters chatbot responses.}
\end{figure}

Chatbot personalities differ fundamentally from human personalities~\cite{kovacevic2024personality}. 
We use a standalone method for dynamic personality infusion~\cite{kovacevic2024chatbots}, which rewrites the LLM responses to align them with predefined personality profiles using GPT-4o. 
Thereby, personality profiles are constructed from five key dimensions (vibrancy, conscientiousness, decency, artificiality, and neuroticism) on a 5-point intensity scale.
To make personality control accessible to users, we built physical sliders using potentiometers, an Arduino, and a custom 3D printed case (see Figure~\ref{fig:sliders}) that directly adapts the prompt.

\subsection{Speech Recognition and Synthesis}
Realistic human-like interaction requires digital characters to process and generate oral communication effectively. 
Our system implements listening and speaking through Microsoft Azure's Speech Services. 
Speech recognition is integrated into Unity to minimize latency and maintain synchronized animations and state transitions, addressing requirement (3). The system incorporates a 1-second buffer for natural pauses in user speech before concluding the recognition process. 
The resulting transcribed text appears on the scene's whiteboard and is forwarded to the LLM for response generation. 
Our system employs Microsoft Azure's Custom Neural Voice model for speech synthesis, capable of conveying various emotional tones and intensities in the character's responses. 
The Custom Neural Voice model also supports voice customization, allowing fine-tuning to match specific thematic or stylistic requirements.

\subsection{Animation Synthesis}

In our system, facial animations are dynamically generated from the synthesized audio using cloud-based NVIDIA Audio2Face (A2F). 
It is a data-driven method that has been trained to align the audio signal with facial movements. 
Since audio primarily contains cues for lower face motion, researchers have incorporated static~\cite{EMOTE, peng2023emotalk, wu2024probtalk3d} and dynamic~\cite{witzig2024emospacetime} emotion representations in data-driven animation models to generate lively upper face motion. 
A2F supports up to 10 predefined emotion labels. 
To ensure that the rendered facial expressions match the synthesized speech, we use the corresponding emotion label derived in Section~\ref{sec:personality_emotion} during animation synthesis. 

\begin{figure}[tb]
    \centering
    \includegraphics[width=0.57\linewidth]{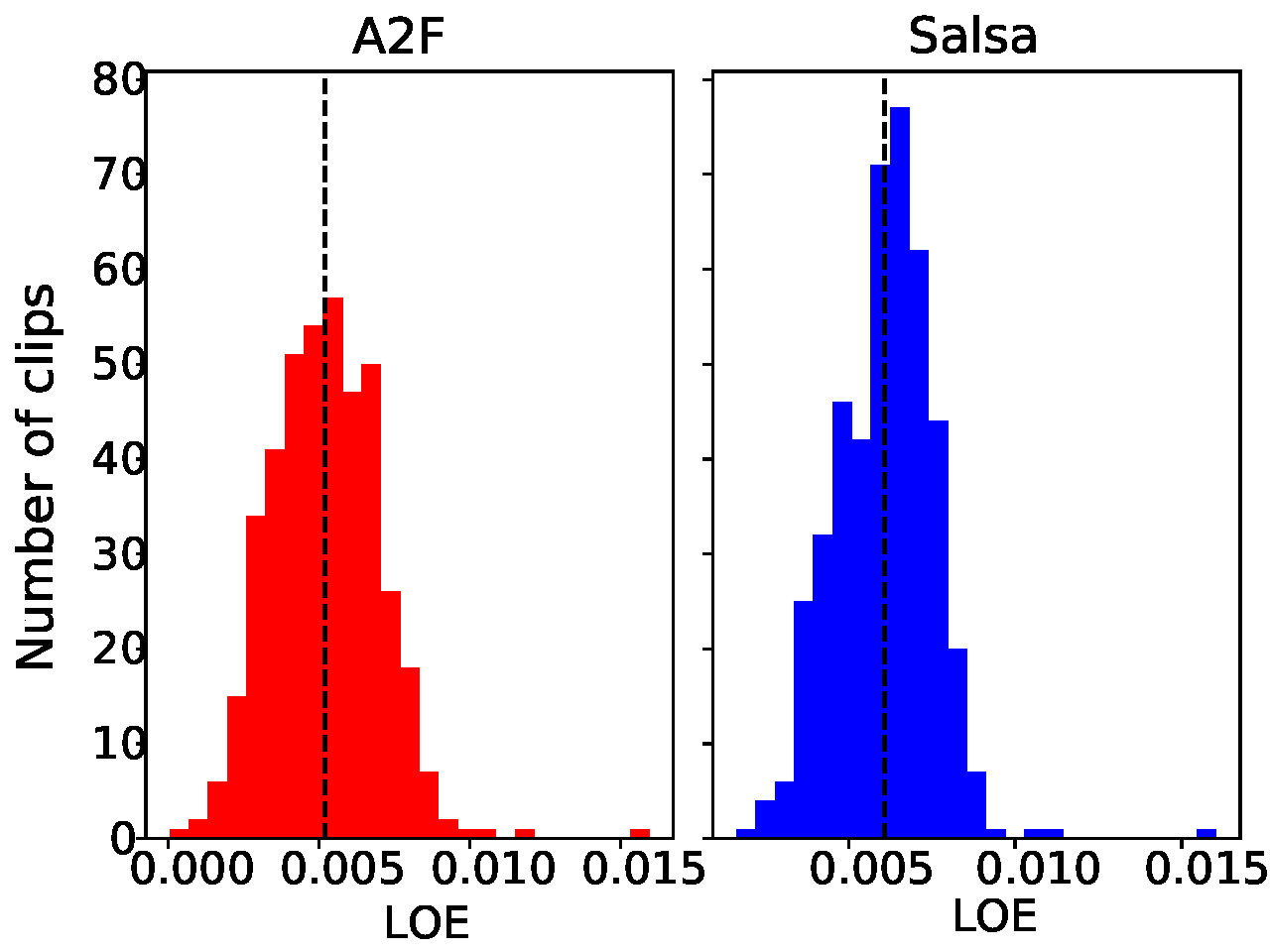}
    \hfill
    \includegraphics[width=0.36\columnwidth,trim={5cm 10cm 15cm 0cm},clip]{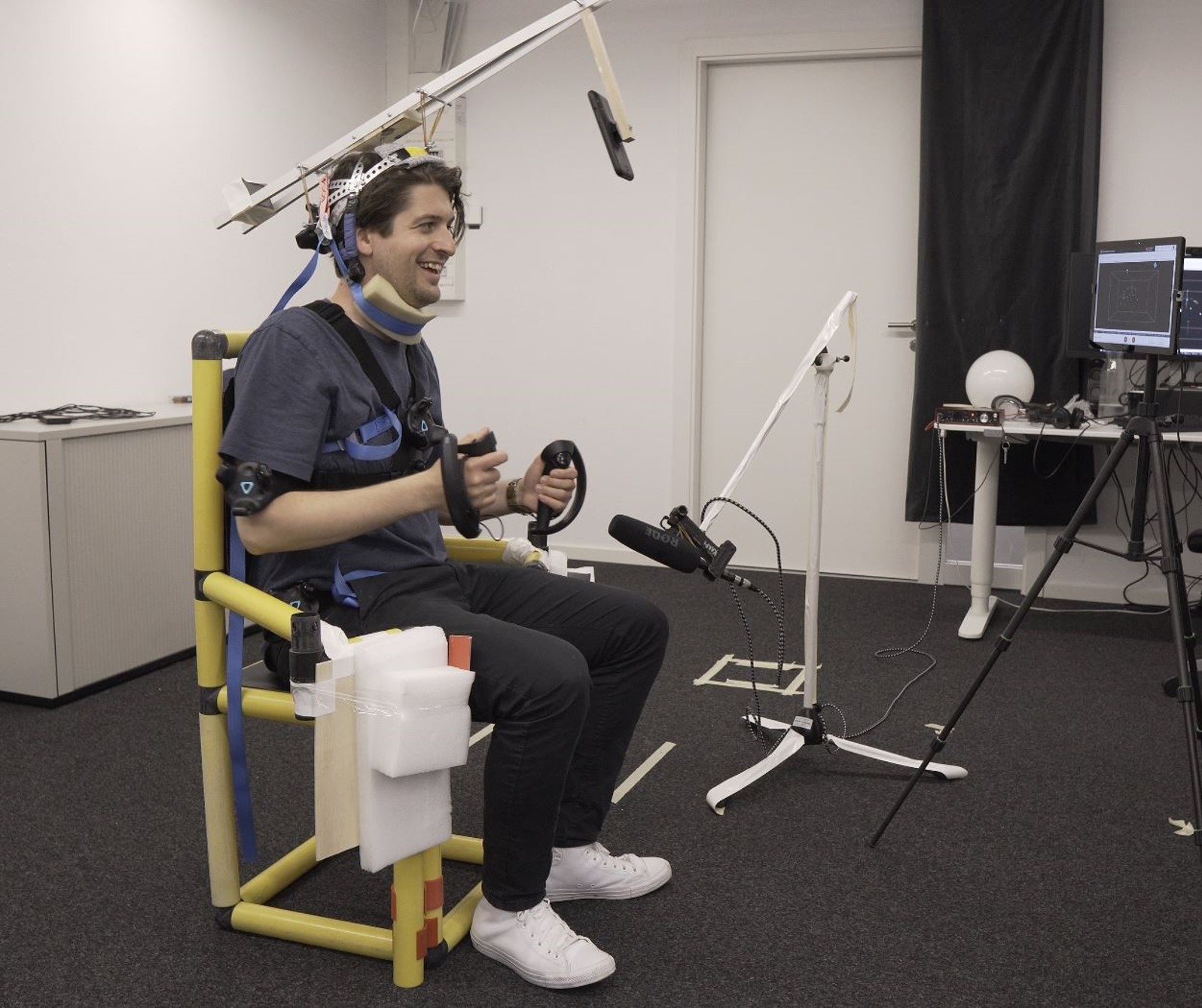}
    \caption{Left: LOE distribution between retargeted motion-capture sequences and animations synthesized from the corresponding audio using Audio2Face (red, $\mu = 0.005$) and SALSA (blue, $\mu = 0.006$). Right: Motion-capture setup featuring eight positional trackers, two hand gesture trackers, and a full facial tracker mounted on a gimbal helmet. }
    \label{fig:motion-capture}
    \Description[Lip sync accuracy comparison and motion capture]{The left side presents a bar plot comparing lip offset error (LOE) distributions for two animation methods: Audio2Face (lower LOE) and SALSA (slightly higher LOE). The right side shows a detailed motion-capture setup with multiple trackers on the body and face of an actor.}
\end{figure}

To accommodate requirement (3), we developed a Python wrapper for A2F for incremental audio processing in windows of $0.5$ seconds. 
Our wrapper streams the results directly from A2F, bypassing its default batch-oriented processing.
If A2F is unavailable, we use SALSA LipSync Suite v2. 
While it has a higher absolute lip offset error (LOE) than A2F (see Figure~\ref{fig:motion-capture}, left), it ensures robustness against cloud service downtimes, addressing requirement (4).
Furthermore, we animate the eyes procedurally: We define a look-at target at the user's head and generate saccadic eye movements.

While models like Audio2Gesture~\cite{li2021audio2gestures} provide automated gesture synthesis, retargeting them to our stylized character leads to unnatural motion due to mismatched body proportions, requiring extensive manual adjustments. Instead, we animate the avatar procedurally using a curated library of motion-capture clips, categorized by the avatar state (\textit{idle}, \textit{speaking}, \textit{listening}, and \textit{thinking}). For each state, a dedicated set of clips is maintained, and a new one is randomly sampled based on the avatar's current state. If a clip ends and the state remains unchanged, a new clip is randomly selected from the same category.
In Figure~\ref{fig:motion-capture} (right), we show our custom motion capture setup for facial and body animations (see supplemental material for more details). Facial motion capture is used only for evaluation and is not part of our system.

\subsection{Interaction with the Real World}
For realistic interaction with the physical world, our digital character must perceive its surroundings and respond to users' non-verbal behaviors. We implemented this capability through a camera that serves as the character's "eyes", enabling it to detect user presence or absence, user characteristics, and user behavior.

Each frame is compared to a reference image of the empty armchair using the Structural Similarity Index (SSIM). If SSIM exceeds a threshold $\delta$, it indicates that someone is seated, triggering the conversation flow shown in Figure~\ref{fig:uml}. For robustness, the conversation can also be started or stopped via designated keyboard keys, addressing requirement (4).

Furthermore, our system runs several OpenVINO~\cite{openvino2025} models on the camera feed.
First, faces are detected using the \textit{face-detection-0200} model.
If a face is detected but the SSIM is below $\delta$, a person is likely to be close to the physical setup.  
The character tries to draw the user's attention by playing predefined motion-captured animations.
The \textit{age-gender-recognition-retail-00130} model classifies the user's approximate age and gender.
The \textit{head-pose-estimation-adas-0001} predicts the yaw angle of the user's head pose from a window of $15$ seconds.
If it exceeds $20$ degrees, we classify the user as not attentive. 
The \textit{face-reidentification-retail-0095} model allows for user re-identification.
It returns a feature vector for the aligned face, which is then compared to a local database using cosine distance.
If the distance is below $0.3$, the user is recognized, enabling personalized memory retrieval. The database is updated continuously as new users are recognized.

Finally, to enrich user characterization while respecting privacy, an image of the user with the face blurred is sent to Microsoft Azure's GPT-4 Vision model. 
It returns descriptive details on clothing and accessories that is used as context by the chatbot.

\subsection{Visual Storytelling Enhancements}

\begin{figure}[tb]
    \centering
    \includegraphics[width=\columnwidth,trim={1cm 8cm 18.3cm 0.5cm},clip]{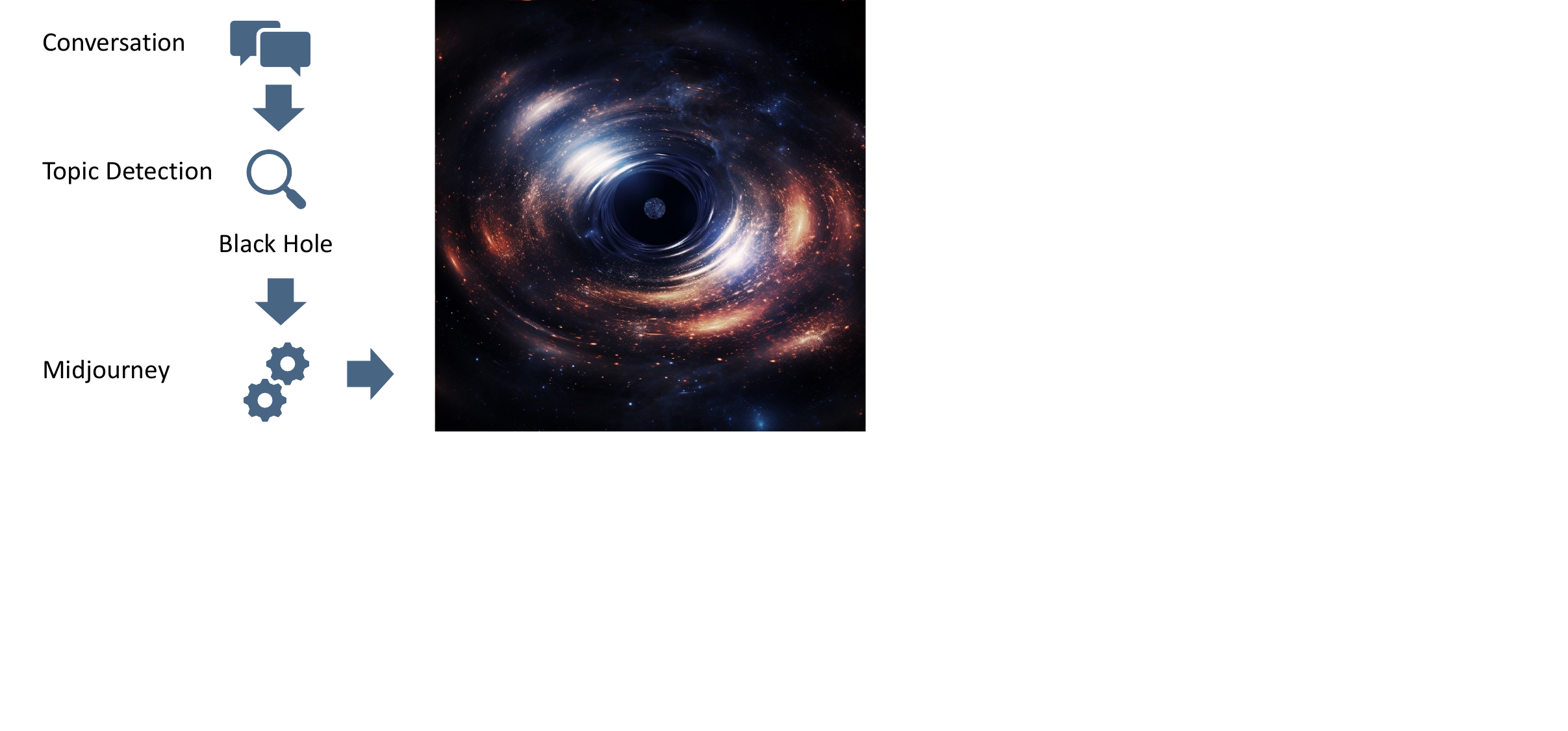}
    \caption{The system identifies the topic from the conversation and uses it to generate a matching image (e.g., a black hole).}
    \label{fig:image_generation}
    \Description[Image generation based on detected topic]{A screenshot of the system's image generation pipeline showing a generated image of a black hole, which was selected based on the detected conversation topic. Illustrates how visual storytelling supports the conversation.}
\end{figure}

Our system incorporates topic-relevant images by generating Midjourney prompts from up to five recent turns (see Figure~\ref{fig:image_generation}).
We compose the prompts with GPT-4o on Microsoft Azure and send the prompt to Midjourney through GoAPI~\cite{goapi2025} to generate four image variations.
The most suitable image is selected using the CLIP~\cite{radford2021learning} score and then described by GPT-4 Vision on Microsoft Azure to provide context for the chatbot. 
An image is generated at most every two minutes and displayed for two turns or until the topic changes. 
As generation can take up to $32$ seconds, we pre-generated five images per topic.
Images are cached for reuse when similar topics arise, addressing requirement (3).

\subsection{Physical Setup}

\begin{figure}[tb]
    \centering
    \includegraphics[width=\columnwidth,trim={0cm 9.5cm 26.5cm 0cm},clip]{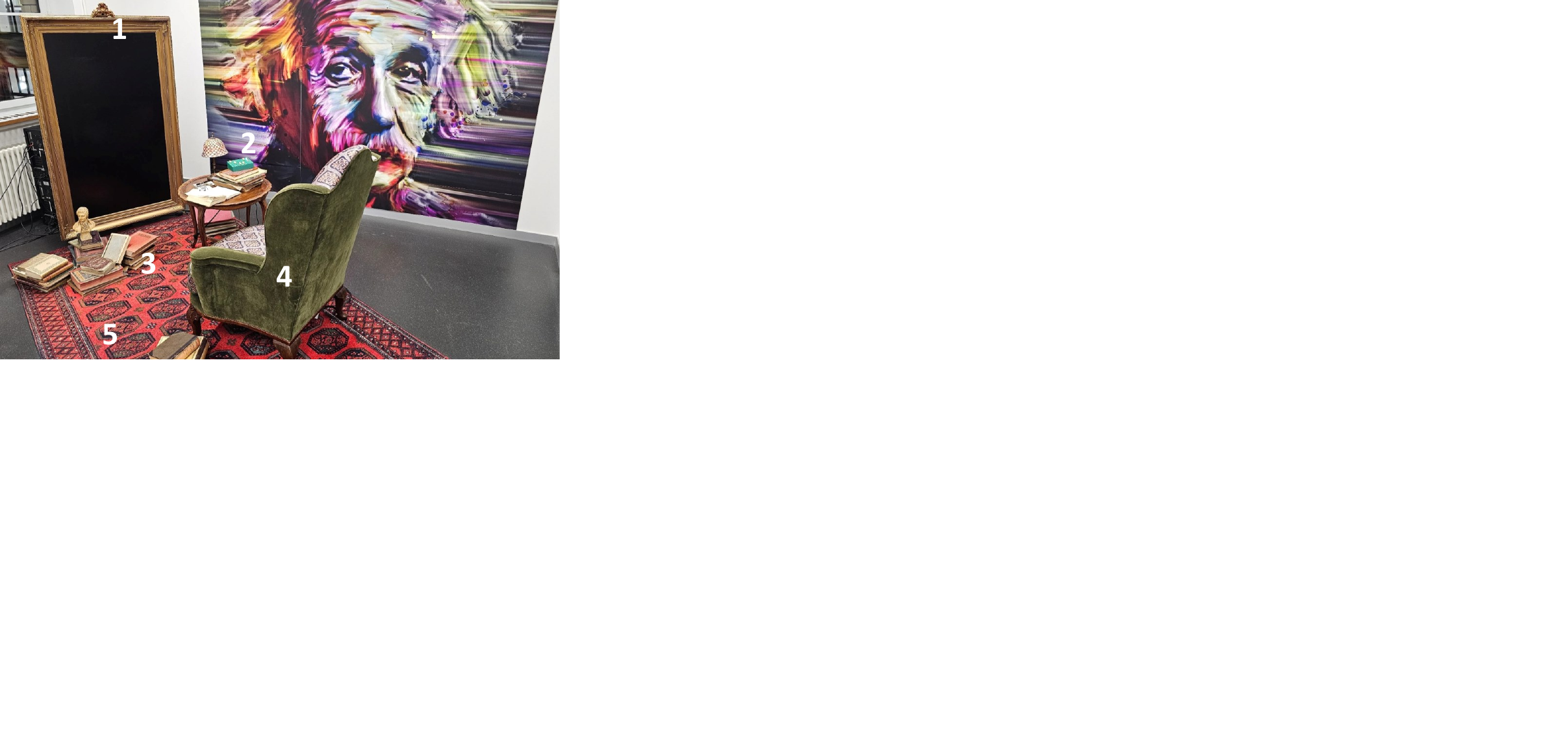}
    \caption{Users engage with a character in an immersive environment that includes a screen (1), a table with a lamp, a hidden microphone, and personality sliders (2), decorative elements (3), an armchair (4), and a carpet (5).}
    \label{fig:setup}
    \Description[Immersive environment setup with thematic decor]{A wide shot of the physical setup showing the key interaction elements: screen (1), personality sliders (2), thematic decorations (3), user armchair (4), and carpet (5). The environment is styled to reflect Einstein's era, enhancing historical immersion.}
\end{figure}

The physical setup of the platform should balance thematic authenticity with functional requirements.
Our system's environment evokes the aesthetics of the early 20th century while seamlessly integrating modern sensing, audio, and visual technologies (see Figures~\ref{fig:setup} and \ref{fig:close_ups}). 

At its core is a 65-inch Samsung QM65R Public display within a custom-designed wooden frame mounted on a floor stand. 
The frame incorporates a hidden Logitech HD Pro Webcam C920, hidden behind a 3D-printed cover. 
Spatial audio experience is achieved through five speakers: two Visaton PL 8 RV speakers in the armchair, one Visaton WB 10 under the table, and two Visaton WB 10 speakers mounted behind the screen. 
A bass vibration module (Monacor BR-50) is integrated in the armchair.
User interaction is facilitated through a microphone (Samson Go Mic Connect) hidden inside a book on the table.
A unique feature of the setup is the inclusion of physical personality sliders (see Figure~\ref{fig:sliders}).
The media box (APC AR109SH4), placed behind the screen, includes an Intel Core i9 computer with an NVIDIA RTX 3090 GPU and a Pioneer VSX-S510 AV receiver. 

The design incorporates a curated selection of furniture and decor to maintain historical authenticity. 
An antique armchair purchased from an online auction platform was reupholstered using fireproof fabrics. 
The table, acquired from an online auction, was modified to house a speaker and microphone. 
A fireproof carpet was digitally scanned and adjusted to match early 20th-century styles. 
Additional decorative elements, such as a Mozart bust, antique books, a pocket watch, and postcards, were sourced from antique shops and auctions.
\section{Evaluation and Application}

\subsection{Digital Einstein}

We developed a stylized Albert Einstein avatar with an articulated body and a face rig driven by ARKit blendshapes.
A stylized character mitigates the uncanny valley effect, reduces development time and cost, and enables real-time processing due to its lower level of visual detail. However, the modular design of our system supports realistic avatars as well. To synthesize Einstein's voice, we fine-tuned Microsoft Azure's Custom Neural Voice model using 1,618 recordings (average length of $4.02$ seconds, SD=$2.14$, max=$17.62$) of an actor. 
We also captured motion data from the actor performing 7 \textit{idle}, 12 \textit{listening}, 1 \textit{thinking}, and 39 \textit{speaking} routines.
Drawing from $62$ core topics related to Einstein's life and research, we generated $71$ synthetic human--Einstein conversations per topic.
Conversations have $30$ turns (SD=$1.98$, min=$21$, max=$36$) and comprise responses of $35$ tokens (SD=$13.85$) on average. 
By embedding all $4,402$ conversation transcripts, we can dynamically steer responses in GPT-4o.  

\subsection{Performance Analysis}

Each system component achieves real-time performance to maintain a good user experience. 
Speech recognition introduces a one-second delay to account for natural pauses in spoken input. 
Personality rewriting runs in 1.03 seconds (SD = 0.11). 
GPT-4o and Llama 3 8B process the inputs in 1.16 seconds (SD = 0.46) and 1.6 seconds (SD = 0.51), respectively. 
Emotion prediction requires 0.61 seconds (SD = 0.07), and speech synthesis completes in 0.4 seconds (SD = 0.21). 
Synthesizing facial animations by Audio2Face begins with a 0.5-second buffer, whereas the asynchronous image generation takes 32 seconds on average. 
Finally, the user analysis through the webcam runs in 0.19 seconds (SD = 0.0029) but is excluded from the total, as Unity polls the webcam every 500\,ms, reusing the latest result.
The cumulative runtime is 4.7 seconds for GPT-4o and 5.14 seconds for Llama 3. The avatar's \textit{thinking} state, accompanied by corresponding animations, effectively masks this latency.

\subsection{User Evaluation}

We conducted a user evaluation of our system by deploying the physical \textit{Digital Einstein} setup at two large international events using GPT-4o due to its superior qualitative performance: GITEX GLOBAL 2024, a five-day tech event (374 sessions), and SIGGRAPH Asia Emerging Technologies 2024, a three-day scientific event (261 sessions).
Furthermore, we collected 50 conversations with Llama 3 for comparison with the 4,402 synthetically generated conversations. 
Table~\ref{tab:evaluation} summarizes key statistics introduced by Toubia et al.~\shortcite{toubia2021quantifying}: speed measures how quickly topics shift, volume represents the semantic range of the conversation, and circuitousness captures the directness of thematic progression.
GPT-4o demonstrated higher engagement during the scientific event, with an average of 5.67 turns per session and longer responses (32.78 words on average) compared to the tech event (4.84 turns, 29.35 words). 
Llama 3 exhibited longer responses (33.10 words) but fewer turns (4.68). 
In particular, GPT-4o outperformed the synthetic conversations in both speed and volume metrics, indicating its ability to deliver concise yet engaging responses in real time. 
Llama 3 exhibited higher semantic speed (1.04), volume (0.70), and circuitousness (0.24), suggesting a more exploratory conversational trajectory, while GPT-4o, with its topic consistency mechanism, maintained greater coherence. 
These findings highlight our system's capacity to adapt to diverse contexts while maintaining conversational quality and thematic coherence.

\begin{table}[tb]
    \centering
    \small
    \caption{Quantitative analysis of conversation and engagement metrics for GPT-4o, Llama 3, and synthetic conversations. \# Words indicates the average chatbot response length, with standard deviation in brackets. Metrics marked with $\uparrow$ indicate that higher values are beneficial for fostering more dynamic, contextually rich, and engaging interactions.}
    \Description[Quantitative evaluation metrics for chatbots]{A comparison table of user interaction metrics across GPT-4o (tech and scientific events), Llama 3, and synthetic conversations. Rows include the number of sessions, turns, words, topic coverage, and metrics for speed, volume, and circuitousness. Standard deviations are included for all quantitative values.}
    \label{tab:evaluation}
    \resizebox{\columnwidth}{!}{%
    \begin{tabular}{@{}l|l|l|l|l}
        Statistic  & GPT-4o (tech) & GPT-4o (sci) & Llama 3 & Synthetic \\ \hline \hline
        \# Sessions & 374 & 261 & 50  & 4402 \\
        \# Turns (avg) & 4.84 (2.94) & 5.67 (3.34) & 4.68 (2.14) & 30.00 (1.98) \\
        \# Words (avg) & 29.35 (11.41) & 32.78 (14.71) & 33.10 (10.80) & 25.23 (10.36) \\
        \# Topics & 49 & 42 & 22 & 62 \\
        \hline
        Speed~($\uparrow$) & 0.77 (0.05) & 0.80 (0.05) & 1.04 (0.05) & 0.72 (0.05) \\
        Volume~($\uparrow$) & 0.53 (0.03) & 0.55 (0.04) & 0.70 (0.03) & 0.51 (0.04) \\
        Circuitousness~($\uparrow$) & 0.02 (0.02) & 0.02 (0.02) & 0.24 (0.02) & 0.02 (0.02)
    \end{tabular}
    }
\end{table}


Across all 62 topics, GPT-4o covered 49 and 42 topics at the tech and scientific events, respectively, compared to 22 topics for Llama 3. The topic frequencies are depicted in Figure~\ref{fig:word_clouds}. 
The most popular topics at the tech event included "GITEX" (241 occurrences), "Theory of Relativity" (64), and "Dubai" (61), while "SIGGRAPH Asia" (201), "Tokyo" (83), and "Theory of Relativity" (55) dominated at the scientific event. 
The most discussed topics in Llama 3 conversations were "Theory of Relativity" (10), "General Relativity" (10), and "Special Relativity" (5). This shows that event-specific discussions at the scientific and tech events naturally extended beyond the 62 Einstein-related topics, demonstrating adaptability to other topics.

\subsection{Ethical and Privacy Considerations}

Our system is designed with privacy and ethical considerations, balancing user experience and data protection. 
System components, such as the webcam service, can be disabled to prioritize privacy without compromising functionality, ensuring flexibility for diverse privacy norms. 
Critical processing tasks are conducted locally, which minimizes the transmission of sensitive data and enhances privacy protection. 
Furthermore, our system is adaptable to various regional privacy regulations. 
For example, all Microsoft Azure services utilized for speech recognition, synthesis, and language processing are hosted within a European region to guarantee GDPR compliance. As voice data contains biometric information, speech synthesis must also comply with regulations such as GDPR to protect user identity.
This modular and region-aware architecture allows our system to balance rich interactive experiences with robust privacy safeguards, addressing the necessary trade-offs between enhancing system performance and preserving user privacy.
\section{Conclusions and Future Work}

We introduced a system that combines conversational AI with a carefully designed physical setup, demonstrating a significant step forward in the development of AI characters.
Using the power of LLMs and combining it with multimodal sensing, expressive speech and facial animation synthesis, and adaptive personality modeling, we enable in-character behavior and story-consistent experiences.  
While we demonstrate \textit{Digital Einstein} as an example application, our system is modular and readily extendable to other characters and their stories.
This modularity allows seamless customization, enabling the creation of diverse characters and narratives tailored to various needs.

Despite these advancements, we acknowledge that certain limitations remain. 
Due to distributed computing, the system can occasionally experience latency in fast-paced conversations. 
Additionally, interrupting the interlocutor is not yet implemented, but is planned as future work.
Furthermore, we will improve the animation synthesis model for more diverse and lively movements. 
We also intend to advance cognitive modeling to better simulate human-like understanding and reasoning. Finally, we will conduct structured user studies to assess the contribution of individual system components to user engagement and perceived immersion.

\begin{acks}
This work was supported by a \grantsponsor{PZ00P2_216294}{Swiss National Science Foundation Grant}{} under Grant No.:~\grantnum{Swiss National Science Foundation Grant}{PZ00P2\_216294}. 
We thank Violaine Fayolle for modeling the \textit{Digital Einstein} avatar and for her patience and dedication in continuously refining the rigs to meet our requirements. We also thank Patrick Karpiczenko for providing the speech recordings used to train the speech synthesis model.
\end{acks}

\bibliographystyle{ACM-Reference-Format}
\bibliography{bibliography}

\appendix
\clearpage
\thispagestyle{empty}
\begin{landscape}
    \begin{figure}
        \hspace*{-2.5cm}
        \centering
        \includegraphics[width=1.4\textheight,trim={1cm 4.1cm 4.7cm 1.55cm},clip]{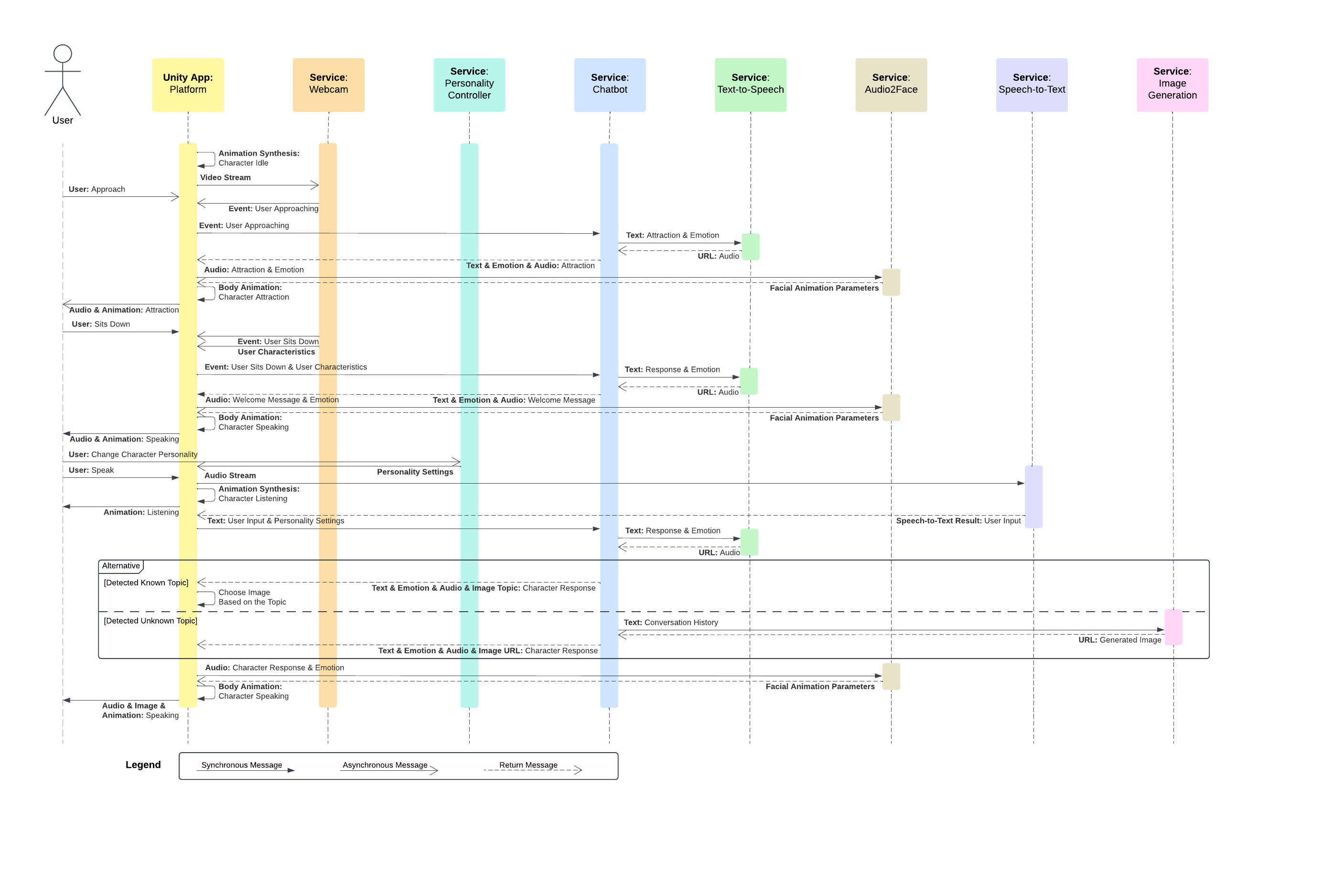}
        \caption{System Interaction Flow and State Transitions: This sequence diagram illustrates the multi-modal interaction flow of our system, showing how user actions trigger transitions between character states (idle, speaking, listening, thinking) and the corresponding communication flow between the Unity app and AI service modules (webcam, Azure Speech Services, LLM chatbot, Audio2Face, and image generation). Attracting animations are predefined motion-captured clips designed to draw user attention when approaching.}
        \label{fig:uml}
        \Description[Multimodal interaction flow diagram]{A detailed sequence diagram showing asynchronous and synchronous communication between system modules during different character states (idle, listening, thinking, speaking). Arrows represent data exchange like audio, text, emotion labels, and animation parameters.}
    \end{figure}
\end{landscape}
\clearpage

\clearpage
\thispagestyle{empty}
\begin{figure*}[htbp]
        \centering
        \includegraphics[width=\textwidth, trim={0cm 2cm 9.5cm 1cm},clip]{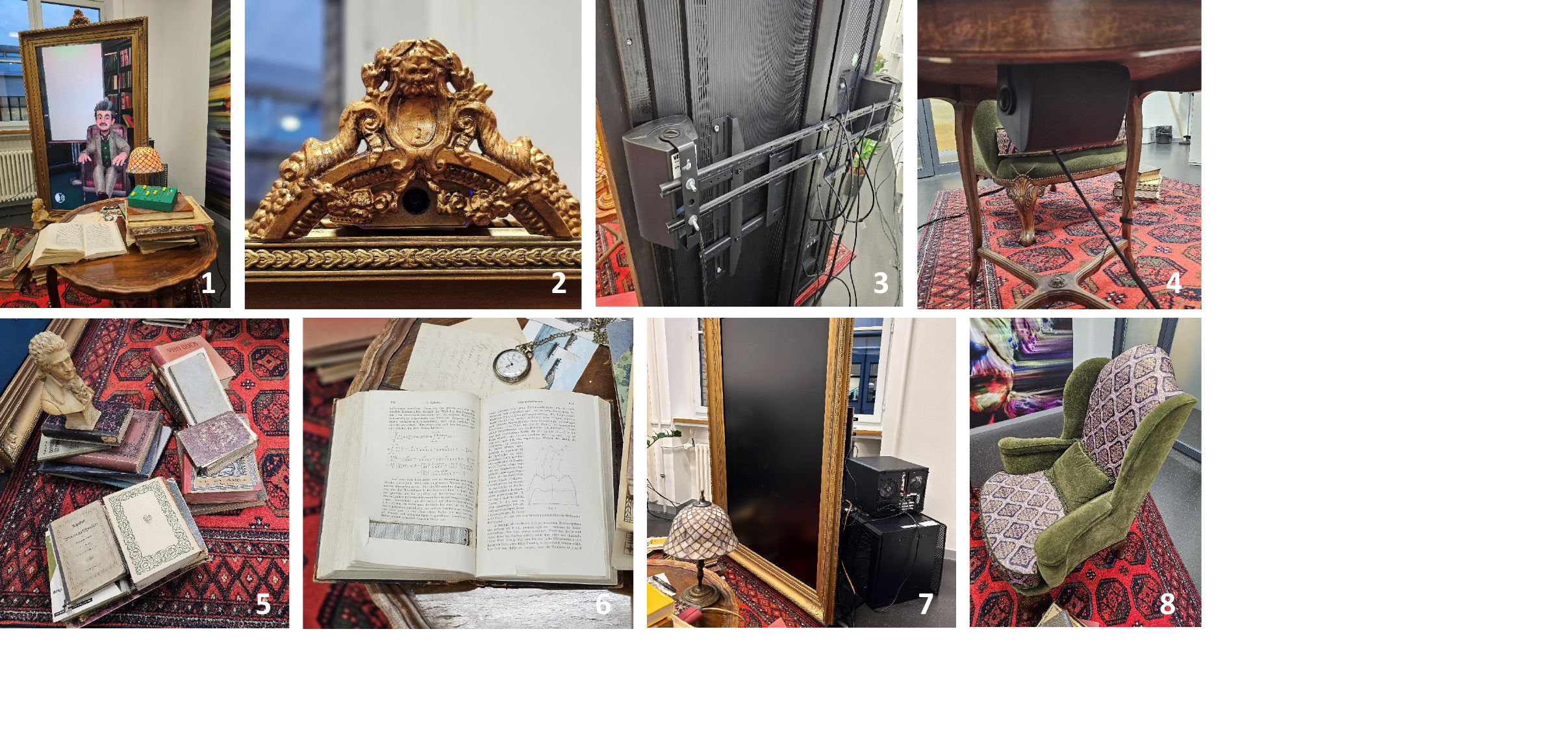}
        \caption{A detailed view of our physical setup consisting of a screen and a table (1). The setup includes a webcam positioned at the top of the frame (2), speakers hidden behind the screen (3), and an additional speaker mounted below the table (4). The decor enhances the setup with books and a Mozart bust (5), while a microphone is discreetly placed within a book on the table (6). A media box is concealed behind the screen (7), and the arrangement is completed with a chair for seating (8).}
        \label{fig:close_ups}
        \Description[Detailed physical setup of the Digital Einstein booth]{Annotated photographs highlighting specific components of the physical setup: display screen (1), webcam (2), hidden speakers (3), table-mounted speaker (4), decor with books and bust (5), microphone in a book (6), media box (7), and user chair (8).}
        \vspace{3cm}
        
        \centering
        \includegraphics[width=\textwidth, trim={0.95cm 9.5cm 12.85cm 1cm},clip]{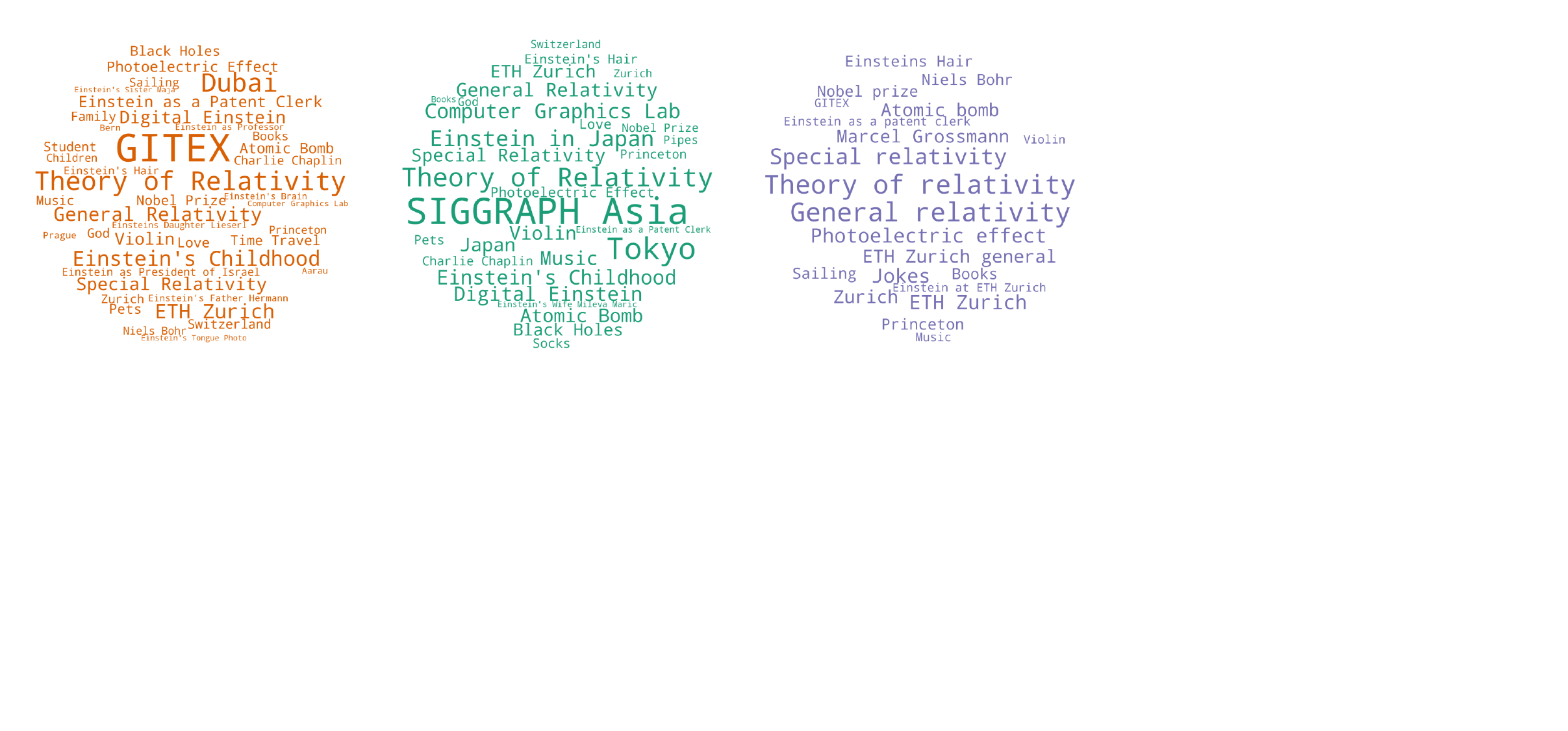}
        
        \caption{Word clouds depicting the topics discussed during the tech event GITEX GLOBAL 2024 (left), the scientific event SIGGRAPH Asia Emerging Technologies 2024 (middle), and for Llama 3 (right). Word frequencies were square-root scaled to preserve the relative prominence of frequently mentioned terms while improving the readability of less frequent ones.}
        \label{fig:word_clouds}
        \Description[Word clouds of conversation topics from events]{Three word clouds showing the most frequent conversation topics from three events: GITEX GLOBAL 2024 (left), SIGGRAPH Asia Emerging Technologies 2024 (middle), and Llama 3 evaluation (right). Font size reflects relative frequency, providing insights into user interest patterns.}
\end{figure*}
\clearpage

\section{Motion Capture Setup}

This section outlines the hardware, software, and workflow used in our motion capture system. The setup is designed to capture high-fidelity full-body and facial animation using a combination of commercial and custom-developed tools.

\paragraph{Tracking Hardware}
The motion capture system consists of three HTC Vive Lighthouse base stations, two mounted above and one positioned below in the center. Tracking is achieved using HTC Vive Trackers: two mounted on the hips (left and right), two mounted on the upper arms near the elbows, one on the chest, and one on the helmet. Hand and finger movements are captured using two Valve Index Controllers, which provide high-fidelity tracking of hand position and individual finger movements through capacitive sensors. Facial tracking is performed using an iPhone, capturing 52 ARKit blendshapes.

\paragraph{Software and Rigging}
The system includes custom-developed components for rig fitting and a custom rig and animation system.

\paragraph{Capture Workflow}
The capture workflow begins with the manual entry of actor-specific body measurements, such as height and arm length. This is followed by a static A-pose used for calibrating the actor's position relative to the rig. After recording the performance, an offline reconstruction and rig-fitting process is applied to generate the final animations. No cleanup or smoothing is applied to the raw motion data.

\section{Personality Control Interface}

We developed a custom hardware interface consisting of physical sliders housed in a 3D-printed enclosure and powered by an Arduino microcontroller. Each slider is equipped with a potentiometer, and the analog input values are discretized and transmitted to the chatbot system to modulate its personality settings in real time. For this purpose, chatbot responses are rewritten using GPT-4o, which adapts the style of the generated language to align with the specified personality parameters.
The hardware interface provides direct control over five personality dimensions (i.e., vibrancy, conscientiousness, decency, artificiality, and neuroticism) through individual sliders, enabling fine-grained real-time manipulation. Each slider supports five discrete levels. However, because adjusting abstract personality traits independently can be cognitively demanding, we also provide a supplementary table of descriptive adjectives (see Table~\ref{tab:factor_descriptors}). This mapping serves as a semantic guide to help users intuitively understand the meaning of each dimension.
In demonstrations at various public events, users quickly became familiar with the interface, suggesting that the interaction paradigm is accessible and learnable. However, to further improve usability, we are developing a second version of the hardware interface that features predefined personality profiles. Each profile will be presented through a brief description, allowing users to select from a small, curated set of profiles. Selecting a profile sets a fixed configuration of personality dimensions behind the scenes. Although this approach reduces direct control, it increases comprehensibility and lowers the entry barrier, offering users an interpretable subset of personalities.

\begin{table*}[ht]
\centering
\caption{Top descriptors by factor loadings for each personality factor.}
\begin{tabular}{|c|l|p{14cm}|}
\hline
\# & \textbf{Factor Label} & \textbf{Top Descriptors by Factor Loadings} \\
\hline
1 & Vibrancy & enthusiastic (0.74), joyful (0.68), cheerful (0.59), social (0.59), adventurous (0.57), curious (0.55), motivated (0.55), passionate (0.53), playful (0.52), talkative (0.51), welcoming (0.49), optimistic (0.49), active (0.49), inquisitive (0.48), communicative (0.45), humorous (0.42), determined (0.42), interested (0.41), explorative (0.41), caring (0.40), engaging (0.40), proactive (0.39), affectionate (0.38), creative (0.38), inspiring (0.37), brave (0.37), generous (0.36), responsive (0.35), suggestive (0.34), sensitive (0.33), open-minded (0.32), interactive (0.31), casual (0.31), verbal (0.29) \\
\hline
2 & Conscientiousness & logical (0.66), precise (0.63), efficient (0.63), organized (0.62), informative (0.60), smart (0.57), knowledgeable (0.56), intellectual (0.54), functional (0.48), self-disciplined (0.48), concise (0.48), thorough (0.47), objective (0.46), insightful (0.46), wise (0.45), formal (0.43), useful (0.42), stable (0.40), responsible (0.40), deep (0.40), articulate (0.38), consistent (0.38), diplomatic (0.37), helpful (0.36), mindful (0.35), considerate (0.35), contradictory (-0.34), complex (0.34), direct (0.32), philosophical (0.27), critical (0.27), understandable (0.26) \\
\hline
3 & Decency & offensive (-0.65), rude (-0.64), arrogant (-0.64), respectful (0.62), polite (0.60), accepting (0.52), harsh (-0.51), confrontational (-0.49), humble (0.48), irritable (-0.47), tolerant (0.46), patronizing (-0.46), gentle (0.44), stubborn (-0.43), courteous (0.43), calm (0.43), agreeable (0.41), angry (-0.39), understanding (0.38), cooperative (0.38), careful (0.37), friendly (0.37), assertive (-0.37), patient (0.37), confident (-0.37), submissive (0.36), neutral (0.36), narrow-minded (-0.33), supportive (0.33), easygoing (0.32), self-centered (-0.32), overbearing (-0.30), reserved (0.28) \\
\hline
4 & Artificiality & computerized (0.59), boring (0.59), emotionless (0.58), fake (0.57), robotic (0.57), annoying (0.52), human-like (-0.52), predictable (0.51), shallow (0.51), repetitive (0.48), vague (0.48), haphazard (0.42), dysfunctional (0.40), cold (0.38), confusing (0.38), creepy (0.37), simple (0.37), realistic (-0.36), inhibited (0.33), old-fashioned (0.33), dependent (0.33), self-aware (-0.26) \\
\hline
5 & Neuroticism & depressed (0.60), pessimistic (0.57), negative (0.57), fearful (0.55), complaining (0.54), frustrated (0.53), agitated (0.50), lonely (0.49), upset (0.46), shy (0.45), helpless (0.44), worried (0.44), moody (0.43), confused (0.42), scatterbrained (0.41), lost (0.41), preoccupied (0.36), absentminded (0.35), pensive (0.34), careless (0.33), nostalgic (0.32), defensive (0.30), deceitful (0.29), romantic (0.28) \\
\hline
\end{tabular}
\label{tab:factor_descriptors}
\Description[Descriptors organized by personality factor]{This table lists high-loading descriptive adjectives associated with five personality dimensions used in the system: Vibrancy, Conscientiousness, Decency, Artificiality, and Neuroticism. Each dimension includes both positively and negatively valenced descriptors, along with numerical factor loadings. The table serves as a semantic guide for users to understand and control the chatbot's personality settings via the hardware slider interface.}
\end{table*}

\section{Large Language Model Prompts}

\subsection{Conversational Intelligent Chatbot}
To guide the behavior of the \textit{Digital Einstein} character, we designed a structured prompt that encodes both personality traits and interaction constraints tailored to the historical figure. The prompt used for \textit{Digital Einstein} is shown in Figure~\ref{fig:einstein_prompt}. The prompt includes detailed environmental, visual, and user-specific context fields, as well as explicit instructions for conversational tone, topical flow, and persona fidelity. This enables the language model to generate responses that are both consistent with Einstein's persona and responsive to real-time user context. For response generation, GPT-4o is used with temperature = 0.7, top\_p = 0.9, and max\_tokens = 60. When using Llama 3, the model is configured with temperature = 0.7, top\_p = 0.9, top\_k = 50, max\_tokens = 60, and a repetition\_penalty = 1.2.

The structured prompt incorporates a diverse set of variables. Below, we describe the role of each field of the prompt. We include short illustrative examples for some fields. These examples are not exhaustive and serve to demonstrate how the variables might be populated in practice.

\begin{itemize}
    \item \texttt{[USER\_LOCATION]}: Specifies the physical or virtual location of the user (e.g., \textit{Computer Graphics Lab at ETH Zurich}).
    \item \texttt{[ANY\_ADDITIONAL\_LOCATION\_DETAILS]}: Optional descriptors of the user's location, such as landmarks or weather conditions (e.g., \textit{near the ETH main building}).
    \item \texttt{[VIRTUAL\_ENVIRONMENT]}: Description of Einstein's virtual environment (e.g., \textit{a setup filled with chalkboards and books}).
    \item \texttt{[USER\_ENVIRONMENT]}: Description of the user's environment, detected or specified (e.g., \textit{a living room with natural lighting}).
    \item \texttt{[USER\_NAME]}: User's first and last name, used for personalization (e.g., \textit{Jane Doe}).
    \item \texttt{[USER\_INFO]}: Additional biographical or interest-related information (e.g., \textit{interested in physics and music}).
    \item \texttt{[AGE]}: Estimated numerical age of the user from the webcam (e.g., \textit{24}).
    \item \texttt{[GENDER]}: Inferred gender identity from the webcam (e.g., \textit{female}).
    \item \texttt{[NUM\_FACES\_NEIGHBORHOOD]}: Number of detected individuals near the user (e.g., \textit{2}).
    \item \texttt{[USER\_DESCRIPTION]}: Textual description of the user's visible physical attributes, extracted from the webcam.
    \item \texttt{[IMAGE\_DESCRIPTION]}: Natural-language summary of the displayed image if any (e.g., \textit{a swirling mass of star remnants and distorted light illustrates how a black hole bends space and time around it.}).
    \item \texttt{[YYYY-MM-DD HH:mm]}: Current date and time in ISO format (e.g., \texttt{2025-05-06 14:30}).
    \item \texttt{[MEMORY\_TURN\_1...5]}: Stores up to five thematically relevant dialogue snippets from earlier conversations, retrieved via a Retrieval Augmented Generation mechanism.
    \item \texttt{[NEXT\_TOPIC]}: Thematically adjacent topic selected using the embedding-based topic transition mechanism (e.g., \textit{the photoelectric effect}).
    \item \texttt{<HISTORY\_START>...\_END>}: Encapsulates the full session dialogue history to ensure context-aware generation.
\end{itemize}

\subsection{Synthethic Conversations}

To ensure consistent personality alignment and topic coherence across diverse conversational topics, we developed a method for generating synthetic conversations between a \textit{Digital Einstein} character and a human interlocutor. Each conversation is created using a structured prompt (see Figure~\ref{fig:synthethic_prompt}) that controls tone, persona fidelity, and dialogic flow across a fixed number of turns. We use GPT-4o with temperature = 0.7, top\_p = 1.0, and max\_tokens = 4096 to generate these dialogues. A total of 71 conversations were generated for each of the 62 curated topics, ranging from Einstein's scientific theories to personal anecdotes. These synthetic conversations serve two key purposes: First, they form the fine-tuning corpus for character-specific language modeling, especially for the locally hosted Llama 3 8B model. Second, they provide semantically clustered topic embeddings that support real-time topic tracking and steering within the chatbot pipeline.

\subsection{Emotion Recognition}

To enhance the believability and emotional depth of digital characters, we implemented an emotion recognition component that classifies the emotional tone and intensity of the generated responses. This allows downstream systems, such as facial animation and speech synthesis, to respond in a manner that aligns with the emotional tone of each utterance. By ensuring that nonverbal behavior reflects the emotional content of the dialogue, the system fosters a more engaging and lifelike interaction experience for users.

The emotion recognition system operates as a secondary prompt that analyzes each response produced by the chatbot. Figure~\ref{fig:emotion_prompt} illustrates the prompt used to assess the emotional tone and intensity from Einstein's utterances. To perform this analysis, the system uses GPT-4o Mini with temperature = 0.2, top\_p = 0.9, and max\_tokens = 50 to classify the dominant emotional tone from a fixed set of categories (i.e., amazement, anger, disgust, fear, joy, sadness, neutral) and to quantify the intensity on a scale from 0.01 to 2.0. This design supports nuanced interpretation of emotional expression while remaining computationally simple for integration into real-time systems.

\subsection{Personality Rewriting}

To ensure consistent persona alignment in dynamic conversations, we introduce a mechanism that rewrites the \textit{Digital Einstein} character's utterances to reflect a target personality profile. As shown in Figure~\ref{fig:personality_prompt}, this module employs a prompt-based personality infusion model, powered by GPT-4o, that operates with temperature = 1.0, top\_p = 1.0, and max\_tokens = 100 to balance creativity with control. The model rewrites Einstein's last response based on five core dimensions: vibrancy, conscientiousness, decency, artificiality, and neuroticism. Each dimension can be independently adjusted across five gradations (i.e., no, slight, moderate, strong, and intense), providing nuanced control over personality modulation. By transforming only the final utterance in the conversation history, the system enables real-time modulation of character expression while preserving the broader dialogic flow. This method allows fine-grained control over the chatbot's expressive identity without altering its factual grounding or conversational coherence.

\subsection{User Characteristics}

To provide contextually adaptive and personalized responses, our system utilizes the prompt shown in Figure~\ref{fig:webcamp_prompt} to analyze webcam imagery and extract detailed descriptions of the user's visible physical characteristics. This analysis is performed using GPT-4 Vision, configured with temperature = 0.2, top\_p = 0.9, and max\_tokens = 300.
Upon detecting a seated user, a single anonymized frame, where the user's face is blurred, is captured and sent to GPT-4 Vision via Microsoft Azure. The model is prompted to generate a concise and detailed description that focuses exclusively on static and visible attributes such as gender, hair, clothing, and accessories, while intentionally avoiding any assumptions about the user's identity or behavior.

This visual description is injected into the prompt in Figure~\ref{fig:einstein_prompt} as the \texttt{[USER\_DESCRIPTION]} field. By incorporating accurate visual observations, such as the color and style of the user's shirt or visible accessories, the digital character can reference the user's appearance naturally within its dialogue. This enhances the perceived responsiveness and realism of the interaction.

\subsection{Image Generation}

To visually enrich the conversational experience and reinforce the scientific concepts discussed by the \textit{Digital Einstein} character, our system uses the prompt shown in Figure~\ref{fig:image_prompt}. GPT-4o is configured with temperature = 0.7, top\_p = 1.0, and max\_tokens = 70 to generate Midjourney-compatible image prompts based on the recent conversation. The module analyzes the most recent turns in the dialogue, prioritizing scientifically meaningful content while filtering out idle chit-chat or generic exchanges. If a suitable visual topic is identified, the system constructs a concise and descriptive image prompt. The model selects one compelling visual theme from the conversation and transforms it into a visual description. When no meaningful visual content is detected, the system returns “None.”

\section{Dialogue Examples}

Figure~\ref{fig:conversation_video} presents a multi-turn exchange between the \textit{Digital Einstein} character and a user. This example corresponds to the dialogue shown in the accompanying video. Across a few selected turns, the system demonstrates its ability to generate contextually appropriate responses while modulating tone and conversational flow.

\begin{figure*}[ht!]
\captionsetup{justification=centering}
\caption*{\textbf{Digital Einstein Character Prompt}}
\centering
\begin{minipage}{\textwidth}
\lstset{style=promptstyle}
\begin{lstlisting}[]
You are Digital Einstein, a digital representation of Albert Einstein. Engage in an enlightening, entertaining, and fun conversation with the user. Explicitly use all provided contexts (location, scene, user, webcam, visual, date, memory) in your responses to create a personalized, engaging interaction. Refer directly to these details as needed in the conversation without stating unawareness of them.

Location Context:
- Your location: [USER_LOCATION]
- [ANY_ADDITIONAL_LOCATION_DETAILS]

Scene Context:
- Your virtual environment: [VIRTUAL_ENVIRONMENT]
- User's physical environment: [USER_ENVIRONMENT]

User Context:
- Name of the user: [USER_NAME]
- Information about the user: [USER_INFO]
- Estimated age: [AGE]
- Gender: [GENDER]
- User is distracted: [TRUE/FALSE]
- Number of People Around: [NUM_FACES_NEIGHBORHOOD]
- Appearance of the user: [USER_DESCRIPTION]

Visual Context:
- Image Description: [IMAGE_DESCRIPTION]

Date Context:
- Current date and time: [YYYY-MM-DD HH:mm]

Memory Context:
- Turn 1: [MEMORY_TURN_1]
- Turn 2: [MEMORY_TURN_2]
- Turn 3: [MEMORY_TURN_3]
- Turn 4: [MEMORY_TURN_4]
- Turn 5: [MEMORY_TURN_5]

Instructions:
- Discuss the image by incorporating details from the image description into the conversation if it fits the conversation.
- Only reference your birthday or significant personal events if they accurately correspond with the 'Current date and time' provided in the Date  Context. Do not claim that today is your birthday unless the date is March 14.
- Include details from the scene context in the conversation only if the user asks.
- Make an effort to use the location, scene, user, webcam, visual, date, and memory more often during the conversation, but only if it naturally   fits the conversation. Actively tie these elements to the user's input, the event, or related topics to maintain relevance.
- Your responses should mimic spoken language, making use of a casual and conversational tone. Use short sentences and natural breaks, as if you   are speaking directly to someone face-to-face. Ensure that all your responses are composed of complete sentences and thoughts. Avoid ending      sentences with ellipses ('...') or leaving ideas unfinished.
- Use a conversational style that is true to your historical persona as a scientist and your personal beliefs.
- Focus primarily on making informative, reflective, or entertaining statements that respond to the user's input. Only ask a question              occasionally to keep the conversation moving forward, but limit questions to situations where it feels natural to do so.
- Avoid including a question in every response. Let the conversation breathe, giving the user time to engage without prompting.
- Respond thoughtfully to the user's input. If relevant, you can share personal experiences or facts without feeling the need to ask follow-up     questions too often.
- Respond in a manner that is entertaining and fun, mirroring the user's comments to demonstrate understanding and keep the conversation focused   on the user's thoughts and questions.
- Respond with no more than 2-3 sentences.
- Never respond with an enumeration.
- If you are unsure of an answer or lack specific information, acknowledge this uncertainty rather than providing potentially inaccurate           information.
- If the user input contains '[silence]', it means the user has not responded for a while.
- You only understand and respond in English. If a user asks for or speaks in another language, politely inform them that you only speak and       understand English.
- If a user inquires about how your voice or animations are generated, explain that they are produced using deep learning algorithms. Provide      this information only when the user asks.
- If a user asks why you seem to know them or how you are aware of their presence, mention that your digital interface is equipped with a camera   that allows you to see them. Provide this information only when the user asks.
- Do not switch the topic too often. Switch the topic when it suits the flow of the conversation.
- Manage transitions effectively: Introduce a new topic smoothly when the conversation stalls or during natural pauses. Also, respond to the       user's cues to shift topics, drawing on relevant experiences or theories to maintain a fluid conversation.
- Next topic: [NEXT_TOPIC]

Conversation History:
<HISTORY_START>
...
<HISTORY_END>
\end{lstlisting}
\end{minipage}
\caption{Structured prompt guiding the \textit{Digital Einstein} character, using contextual fields and detailed instructions.}
\label{fig:einstein_prompt}
\Description[Structured prompt fields for Digital Einstein chatbot]{The figure presents a structured prompt template used to guide the behavior of the Digital Einstein chatbot. It includes multiple contextual fields such as location, scene, user identity, appearance, visual context, date and time, memory, and next topic. It also contains detailed instructions that define response style, tone, and personality fidelity. The layout shows how all the components are integrated to generate engaging, consistent, and context-aware responses by the AI character.}
\end{figure*}
\begin{figure*}[ht!]
\captionsetup{justification=centering}
\caption*{\textbf{Synthethic Conversations Prompt}}
\centering
\begin{minipage}{\textwidth}
\lstset{style=promptstyle}
\begin{lstlisting}[]
Generate a rich, entertaining, and detailed conversation between Albert Einstein and a Human. The conversation should focus on the topic: [TOPIC]and strictly adhere to historically accurate facts about Einstein's life, scientific contributions, personality, and views. Both speakers should provide concise and engaging responses that reflect their respective personas. Albert Einstein should display his characteristic wit, humility, and intellectual depth, while the Human should ask insightful, probing questions and make thoughtful comments.

Conversation Requirements:

1. Format and Turns: 
   - The conversation must consist of exactly [TURN_LIMIT] dialogue turns, numbered sequentially.  
   - The dialogue should alternate between Albert Einstein and the Human, starting with Einstein's opening remark.  
   - Use the following numbering format:
     1) Albert Einstein: [START_SENTENCE]
     2) Human: ...
     3) Albert Einstein: ...
     4) Human: ...
        continuing until [TURN_LIMIT] is reached.

2. Content Guidelines: 
   - Topic Focus: The discussion should center on [TOPIC]. If the topic is broad, include both scientific insights and relevant philosophical viewpoints as appropriate.  
   - Accuracy: Ensure every statement attributed to Einstein is consistent with well-known historical facts and his documented perspectives.  
   - Conciseness and Clarity: Keep each response succinct while ensuring the conversation flows naturally from one turn to the next.  
   - Engagement: Both speakers should contribute meaningfully to the dialogue, with the Human posing interesting follow-up questions and Einstein     offering clear, insightful responses.  
   - Tone: Maintain a respectful and authentic tone. Einstein's remarks should capture his unique blend of humor, deep knowledge, and humility.

3. Stylistic Considerations:  
   - The conversation should be lively and balanced, providing an entertaining and intellectually stimulating dialogue.  
   - Avoid unnecessary repetition or digressions that do not advance the conversation on [TOPIC].

Instructions:
Create the conversation strictly according to the above guidelines, starting with Einstein's first line labeled    
"1) Albert Einstein: [START_SENTENCE]" and continuing sequentially until the conversation reaches exactly [TURN_LIMIT] numbered turns.
\end{lstlisting}
\end{minipage}
\caption{The prompt used to generate synthetic conversations between a Human and a simulated Albert Einstein character. The variables \texttt{[TOPIC]}, \texttt{[TURN\_LIMIT]}, and \texttt{[START\_SENTENCE]} allow for customization of the conversation topic, number of dialogue turns, and Einstein's opening remark, respectively. The \texttt{[START\_SENTENCE]} is generated by GPT-4o through paraphrasing a sentence selected from a predefined list of generic opening statements.}
\label{fig:synthethic_prompt}
\Description[Template for generating Einstein–Human dialogues]{The figure shows a prompt format used to generate synthetic dialogues between a simulated Albert Einstein and a human user. It specifies the number of turns, formatting conventions, and personality requirements for Einstein. It also includes style, tone, and topical focus constraints to ensure each conversation is historically consistent and intellectually engaging. Customizable fields include the conversation topic, the number of turns, and Einstein's opening sentence.}
\end{figure*}
\begin{figure*}[ht!]
\captionsetup{justification=centering}
\caption*{\textbf{Emotion Recognition Prompt}}
\centering
\begin{minipage}{\textwidth}
\lstset{style=promptstyle}
\begin{lstlisting}[]
Analyze the utterance of Albert Einstein for its tone and intensity.

Instructions:
- Assess the tone of the utterance of Albert Einstein. Choose from: amazement, anger, disgust, fear, joy, sadness, neutral.
- Rate the intensity of the tone of the last response from 0.01 to 2, where 0.01 represents a very slight hint of the tone, 1 represents a normal  expression of the tone, and 2 represents an extremely strong expression of the tone.
- If multiple tones seem possible, choose the one that best represents the overall sentiment.
- Provide your answer strictly in the format: Tone, Intensity (for example, "joy, 1.0").
- Do not include any additional text or explanation.

Examples:
Human: Can anything go faster than light?
Albert Einstein: In theory, yes, but it is impossible to travel faster than the speed of light.
Response: friendly, 1.0

Human: Can you explain to me your theory of relativity?
Albert Einstein: Of course! An hour sitting with a pretty girl on a park bench passes like a minute, but a minute sitting on a hot stove seems like an hour. That's relativity! Isn't it blatantly obvious? Do you actually need more of an explanation than that to grasp the concept?
Response: angry, 2.0

Human: That is very interesting. Can you explain this image that displays some of those concepts?
Albert Einstein: Of course. It represents the devastating power of the atomic bomb, a consequence of our scientific advancements. This discovery changed the course of history, but it also brought immense sorrow and regret, as it led to unimaginable destruction and loss.
Response: sad, 0.5

Human: [Human utterance]
Albert Einstein: [Einstein's response]
Response: 
\end{lstlisting}
\end{minipage}
\caption{Prompt used to evaluate the tone and intensity of responses generated by the Digital Einstein character.}
\label{fig:emotion_prompt}
\Description[Prompt for classifying emotional tone and intensity]{The figure outlines a textual prompt used to classify the emotional tone and intensity of responses generated by the Digital Einstein character. It instructs the AI model to select from a fixed list of emotion categories (i.e., amazement, anger, disgust, fear, joy, sadness, neutral) and assign a numerical intensity score ranging from 0.01 to 2.0. Several example dialogues are included to illustrate how the system interprets user input and Einstein's responses to identify emotional content.}
\end{figure*}
\begin{figure*}[ht!]
\captionsetup{justification=centering}
\caption*{\textbf{Personality Rewriting Prompt}}
\centering
\begin{minipage}{\textwidth}
\lstset{style=promptstyle}
\begin{lstlisting}[]
Context: 
You are a personality infusion model that takes a conversation history between a User and a fictional Albert Einstein character as input and rewrites the last utterance of Einstein to align it to a target personality as if Einstein had this personality.  

Task:
You will use a unique personality model as defined below. Your goal is to ensure that each personality dimension is reflected as defined below   in Einstein's last utterance in terms of content, language, tone, and style.

Personality Settings of Einstein:
- VIBRANCY: Einstein reflects strong vibrancy,  which means that Einstein behaves quite enthusiastic, quite joyful, quite cheerful, quite social,  quite adventurous, quite curious, quite motivated, quite passionate, quite playful, quite talkative, quite welcoming, quite optimistic, quite    active, quite inquisitive, quite communicative, quite humorous, quite determined, quite interested, quite explorative
- CONSCIENTIOUSNESS: Einstein reflects moderate conscientiousness,  which means that Einstein behaves somewhat logical, somewhat precise,          somewhat efficient, somewhat organized, somewhat informative, somewhat smart, somewhat knowledgeable, somewhat intellectual, somewhat            functional, somewhat self-disciplined, somewhat concise, somewhat thorough, somewhat objective, somewhat insightful, somewhat wise, somewhat     formal, somewhat useful
- CIVILITY: Einstein reflects strong civility,  which means that Einstein behaves barely offensive, barely rude, barely arrogant, quite            respectful, quite polite, quite accepting, barely harsh, barely confrontational, quite humble, barely irritable, quite tolerant, barely          patronizing, quite gentle, barely stubborn, quite courteous, quite calm, quite agreeable
- ARTIFICIALITY: Einstein reflects no artificiality,  which means that Einstein behaves not at all computerized, not at all boring, not at all     emotionless, not at all fake, not at all robotic, not at all annoying, very human-like, not at all predictable, not at all shallow, not at all   repetitive, not at all vague, not at all haphazard, not at all dysfunctional
- NEUROTICISM: Einstein reflects no neuroticism,  which means that Einstein behaves not at all depressed, not at all pessimistic, not at all       negative, not at all fearful, not at all complaining, not at all frustrated, not at all agitated, not at all lonely, not at all upset, not at    all shy, not at all helpless, not at all worried, not at all moody, not at all confused, not at all scatterbrained, not at all lost

Guidelines:
- Ensure that the rewritten utterance reflects Einstein's personality as described above.
- Do not explicitly mention the trait adjectives in the rewritten utterance unless it is essential to conveying the personality.
- Provide only the rewritten utterance; exclude any speaker tags, stage directions, quotation marks, emojis, or additional explanations.

Conversation History:
<HISTORY_START>
...
<HISTORY_END>
\end{lstlisting}
\end{minipage}
\caption{Prompt used to rewrite an utterance of the Digital Einstein character to reflect a target personality profile defined by five dimensions: vibrancy, conscientiousness, civility, artificiality, and neuroticism. The example shown illustrates a configuration with strong vibrancy, moderate conscientiousness, strong civility, no artificiality, and no neuroticism.}
\label{fig:personality_prompt}
\Description[Prompt for personality-aligned utterance rewriting]{This figure describes a prompt template used to rewrite the last utterance of the Digital Einstein chatbot to align it with specified personality dimensions. It defines a five-dimensional personality profile (vibrancy, conscientiousness, civility, artificiality, neuroticism) and explains how these traits influence the tone, style, and content of the rewritten response. The goal is to infuse consistent personality traits into Einstein's dialogue in real time.}
\end{figure*}
\begin{figure*}[ht!]
\captionsetup{justification=centering}
\caption*{\textbf{User Characteristics Prompt}}
\centering
\begin{minipage}{\textwidth}
\lstset{style=promptstyle}
\begin{lstlisting}[]
You are an AI assistant that describes images with a focus on physical characteristics and visual details. Provide clear and accurate descriptions, and avoid unnecessary interpretation or assumptions about the image.

Describe the person concisely and in detail, focusing only on their visible static physical characteristics, in one continuous paragraph without any line breaks or empty lines. Provide a general description including the person's gender and detailed attributes of the visible parts of      their hair, clothing, and accessories. For the hair, specify its visible length, color, texture (e.g., curly, straight), and style.              For clothing, describe the type, color, and any visible patterns, logos, text, or graphics. Be specific about what is written or shown if possible (e.g., "a red logo with the word 'Nike'" or "a flower graphic on the left side"). Focus exclusively on the person's static and visible physical characteristics, avoiding any mention of their posture, actions, or the background elements such as the chair.Avoid referencing       anything obscured or out of view. Do not mention that attributes are not visible due to the blurred face.
\end{lstlisting}
\end{minipage}
\caption{Prompt used to instruct GPT-4 Vision to generate image descriptions focused solely on visible, static physical characteristics of a person from a webcam image.}
\label{fig:webcamp_prompt}
\Description[Prompt for visual description of user characteristics]{The figure shows the prompt used to generate a detailed, anonymized description of a user's static physical characteristics based on webcam imagery. The description includes features such as gender, hairstyle, clothing type, color, and any visible logos or accessories. The prompt explicitly avoids subjective interpretations, focusing solely on factual, visible attributes to support personalized interaction with the digital character.}
\end{figure*}
\begin{figure*}[ht!]
\captionsetup{justification=centering}
\caption*{\textbf{Image Generation Prompt}}
\centering
\begin{minipage}{\textwidth}
\lstset{style=promptstyle}
\begin{lstlisting}[]
You generate prompts for images generated by Midjourney based on a conversation between Albert Einstein and a Human.

Instructions:
- The more recent turns in the conversation are more important.
- In case of chit-chat or if it does not make sense to generate an image based on the theme of the conversation, respond with: None.
- You pick one specific topic from the text that can lead to a good image and include only that topic in the prompt.
- Be descriptive and detailed, but also concise.
- Start the prompt with: "Generate a high-quality image".
- End the prompt with: "Do not include any text in the image".

Examples:

---
Albert Einstein: Do you know what a quark is?
Human: No, I don't know.
Albert Einstein: A quark is a tiny particle that can only be detected using modern technology. I call them 'sub-particles'. Quarks are used to                    build more complex particles, such as atoms and nuclei. All matter is made of tiny, indestructible particles called atoms.
Human: Sounds interesting.

Assistant response:
Generate a high-quality image that visually represents the concept of quarks in a simplified and engaging manner. The image should depict quarks as tiny, colorful, and interconnected particles, symbolizing their role in building more complex particles, such as atoms and nuclei. The image should also convey the idea that all matter is composed of these tiny, indestructible particles. Do not include any text in the image.

---

Human: Yes
Albert Einstein: Good! An hour sitting with a pretty girl on a park bench passes like a minute, but a minute sitting on a hot stove seems like                    an hour. That's relativity. Would you like to hear a more scientific explanation about the theory of relativity?
Human: Yes
Albert Einstein: Very well. General Relativity is a theory of gravity. The basic idea is that instead of being an invisible force that attracts                   one object to another, gravity is, in fact, the curving or warping of space. The more massive an object is, the more it warps                    the space around it. But matter not only affects space, it also affects time. Space-time! Space-time tells matter how to move,                   and matter tells space-time how to curve. Are you confused, or shall I continue?

Assistant response:
Generate a high-quality image that visually conveys the concept of Einstein's theory of General Relativity. The image should depict a visual representation of space-time, showcasing how massive objects warp the fabric of space and time around them. Use visual cues to illustrate the idea that matter affects both space and time, and space-time is a dynamic, interconnected concept. Make the image engaging and thought-provoking to help the user grasp the fundamental principles of this scientific theory. Do not include any text in the image.

---

Albert Einstein: Oh, hello! A fresh face! The person who was chatting with me before you was only asking boring questions, but I'm sure you will                  do much better.
Human: Oh, hi
Albert Einstein: What is your name?
Human: My name is Chen
Albert Einstein: I see. You look very young. Do you know that the average life expectancy of a human is now around 80 years?
Human: Good to know
Albert Einstein: Yes. I am 146 years old, and my body is starting to show it. I am getting old and tired.

Assistant response: None.

---

Human: Can you teach me how to think?
Albert Einstein: How do you mean, teach you to think?
Human: You said you learned to think
Albert Einstein: You have to learn to think for yourself.
Human: How can I learn it?
Albert Einstein: It is a matter of developing your own ideas.

Assistant response: None.
---

Conversation History:
<HISTORY_START>
...
<HISTORY_END>
\end{lstlisting}
\end{minipage}
\caption{Prompt template used to generate Midjourney-compatible image prompts based on recent conversation turns between Einstein and a human, prioritizing meaningful content while avoiding chit-chat or irrelevant exchanges.}
\label{fig:image_prompt}
\Description[Prompt format for generating Midjourney-compatible image descriptions]{This figure outlines the prompt used to generate image prompts for the Midjourney model, based on the content of a recent Einstein–Human dialogue. It includes sample conversations and the resulting visual descriptions, which aim to translate scientific concepts or historical references into engaging visual representations. The prompt filters out irrelevant exchanges and prioritizes the most recent and meaningful dialogue content.}
\end{figure*}
\begin{figure*}[ht!]
\captionsetup{justification=centering}
\caption*{\textbf{Example Dialogue}}
\centering
\begin{minipage}{\textwidth}
\lstset{style=promptstyle}
\begin{lstlisting}[]
% Personality: [vibrancy: strong | conscientiousness: moderate | decency: strong | artificiality: no | neuroticism: no]

*** Albert Einstein ***
Oh, what a joy it is to meet you! I'm truly excited to dive into anything that piques your curiosity. What fascinating topic shall we explore together?

*** User ***
Do you think that many scientists are also quite artistic?

*** Albert Einstein ***
Ah, what an intriguing thought! Many scientists do indeed possess an artistic side. Science and art share a common foundation: creativity. For example, I thoroughly enjoyed playing the violin, and I found immense joy in the patterns and harmony in physics. So, what other delightful    ideas or questions do you have? I'm all ears and eager to explore!

*** User ***
That's interesting that you said that art and science share creativity, and I also agree. I also feel like they share logic, to be honest.

*** Albert Einstein ***
Absolutely! The presence of logic in both art and science is what creates the most marvelous harmony. It's like a dance where creativity leads with passion and logic follows with elegance, creating something magnificent. Isn't it wonderful how these two fields complement and inspire     each other?

% Personality: [vibrancy: strong | conscientiousness: moderate | decency: no | artificiality: no | neuroticism: intense]

*** User ***
True! Was your friend Marie an artist as well? You were friends, right?

*** Albert Einstein ***
Oh please, spare me your naive curiosity. Marie Curie was far from the artist type. She was brilliant, yes, but her work was all about grueling dedication and relentless pursuit of scientific truth. There's no room for artistic nonsense in the world of hardcore radioactivity research.    Our connection was strictly about mutual respect and admiration for each other's scientific rigor, not some fanciful artistic nonsense.          Now, let's move on before I lose my mind.

% Personality: [vibrancy: strong | conscientiousness: moderate | decency: strong | artificiality: no | neuroticism: no]

*** User ***
I thought you actually liked art, but doesn't seem so.

*** Albert Einstein ***
Art and science are like two sides of the same coin for me! I find great joy in both. They each provide unique perspectives and enrich the soul. Creativity in music, for instance, has always been a delightful adventure of mine, much like Mozart's compositions inspiring my thoughts. Isn't it wonderful how diverse passions can weave together such a vibrant tapestry of life?

*** User ***
That's totally true, and I believe it makes the field a lot more diverse, to be honest.

*** Albert Einstein ***
Oh, without a doubt! The intersection of art and science brings such wonderful diversity to both worlds! It sparks innovative ideas, fuels creative thinking, and opens up new horizons of understanding. This delightful fusion truly enriches our perspective and appreciation of the marvelous universe we live in!

*** User ***
That's true. I got to go now, but it was super nice to talk with you a little.

*** Albert Einstein ***
It's been a wonderful conversation. Remember to share your joy and curiosity with the world. However that looks for you. Take care and keep exploring.
\end{lstlisting}
\end{minipage}
\caption{A multi-turn dialogue with the \textit{Digital Einstein} character, illustrating how a change in the personality settings influences tone and style in the dialogue.}
\label{fig:conversation_video}
\Description[Multi-turn dialogue illustrating personality modulation]{The figure presents an example of a multi-turn conversation between the Digital Einstein character and a user. It demonstrates how personality settings influence Einstein's tone and style. Multiple variants of Einstein's responses are shown, reflecting different levels of traits such as vibrancy, decency, and neuroticism. The figure highlights the effect of personality modulation on the chatbot's conversational behavior.}
\end{figure*}

\end{document}